\documentclass[prd, twocolumn, nofootinbib, floatfix, a4paper]{revtex4}
\usepackage{amsmath}
\usepackage{graphicx}
\usepackage{dcolumn}
\usepackage{bm}
\usepackage{epsfig}
\usepackage{amssymb,latexsym,mathrsfs}
\usepackage{graphicx}
\usepackage{color}

\newcommand{\mnras}{{Mon.~Not.~R.~Astron.~Soc.}}
\newcommand{\be}{\begin{equation}}
\newcommand{\ee}{\end{equation}}
\newcommand{\bea}{\begin{eqnarray}}
\newcommand{\eea}{\end{eqnarray}}

\newcommand{\od}{\Omega_{\rm de}}
\newcommand{\omegae}{\Omega_{\rm e}}

\newcommand{\cvis}{c_{\rm vis}}
\newcommand{\cs}{c_{\rm s}}

\begin{document}
\title{Future CMB Constraints on Early, Cold, or Stressed Dark Energy}

\author{Erminia Calabrese$^1$, Roland de Putter$^{2,3,4}$, Dragan Huterer$^5$, 
Eric V.\ Linder$^{4,6}$, Alessandro Melchiorri$^1$} 
\affiliation{$^1$ Physics Department and INFN, Universita' di Roma ``La Sapienza'', Ple Aldo Moro 2, 00185, Rome, Italy.}
\affiliation{$^2$ Instituto de Fisica Corpuscular, Valencia, Spain}
\affiliation{$^3$ Institut de Ciencies del Cosmos, Barcelona, Spain}
\affiliation{$^4$ Berkeley Lab \& University of California, Berkeley, CA 94720, USA.} 
\affiliation{$^5$ Department of Physics, University of Michigan, 450 Church St, Ann Arbor, MI 48109, USA.}
\affiliation{$^6$ Institute for the Early Universe, Ewha Womans University, Seoul, Korea}

\begin{abstract}
We investigate future constraints on early dark energy (EDE) achievable by the
Planck and CMBPol experiments, including cosmic microwave background (CMB)
lensing.  For the dark energy, we include the possibility of clustering
through a sound speed $\cs^2<1$ (cold dark energy) and anisotropic stresses
parameterized with a viscosity parameter $\cvis^2$.  We discuss the
degeneracies between cosmological parameters and EDE parameters. In particular
we show that the presence of anisotropic stresses in EDE models can
substantially undermine the determination of the EDE sound speed parameter
$\cs^2$. The constraints on EDE primordial energy density are however
unaffected. We also calculate the future CMB constraints on neutrino masses
and find that they are weakened by a factor of 2 when allowing for the
presence of EDE, and highly biased if it is incorrectly ignored. 
\end{abstract}

\maketitle

\section{Introduction}
\label{sec:intro}

For about a decade cosmological data from cosmic microwave background 
(CMB) anisotropy experiments (\cite{wmap7}, \cite{acbar}, \cite{quad}), 
in combination with complementary results from galaxy surveys \cite{2dF,SDSS} 
and Type Ia supernovae \cite{PerlRiess,union2}, suggest in an unequivocable 
way that the present energy budget of the universe 
is dominated by an exotic form of energy coined {\it dark energy\/}.

The presence of a cosmological constant term $\Lambda$ in Einstein's equation
of General Relativity (GR) is the simplest explanation for dark energy.  The
Lambda cold dark matter scenario ($\Lambda$CDM) is a simple model that
consistently accounts for all observations, and has therefore emerged as the
standard model of cosmology.  Despite the simplicity of this concordance
model, however, the presence of a tiny but nonzero cosmological constant is
vexing, and is not understood from the point of view of fundamental theory 
(see e.g.\ \cite{lambda} and references therein).  Dark energy could therefore
be different from a cosmological constant, and indeed many diverse models are
also consistent with the data \cite{beylam,sollerman,mhh2}.

Within the framework of a non-interacting, minimally coupled additional
component to the energy density, a general dark energy fluid and the
cosmological constant may differ in two main aspects: the latter behaves as a
homogeneous fluid with a constant energy density, while the former is a
non-homogeneous fluid with a time dependent energy density and pressure.  A
simple way of describing these models is by specifying the equation of state
$w=p/\rho$, where $p$ and $\rho$ are the dark energy pressure and density.
The cosmological constant corresponds to $w=-1$, while a general dark energy
fluid may have a time dependent equation of state $w(a)$ which is as function
of the scale factor $a(t)$, so that $w\neq-1$ in general.

Density perturbations in the dark energy component 
could also leave an imprint in cosmological observables, 
while $\Lambda$ is purely homogeneous. 
The clustering properties of different dark energy models are usually
parameterized by an effective sound speed,
defined as the ratio between the pressure to density perturbations in the rest
frame of dark energy; $\cs^2=\delta p/\delta \rho$ (see, e.g.,
\cite{Hu98,Bean,JochenLewis}).
Moreover, anisotropic stress can also affect the density perturbations. For
example, in the case of a relativistic component, anisotropic stresses act as
a form of viscosity in the fluid and damp density pertubations. If dark energy
behaves like a relativistic fluid in the past, then the effects of viscosity
should also be considered.

To parameterize viscosity in a dark component one can introduce 
the viscous sound speed $\cvis^2$, which controls the
relationship between velocity/metric shear and the anisotropic stress
\cite{Hu98,Koivisto:2008ig,mota}. A value of $\cvis^2=1/3$, for example, is
what one expects for a relativistic component, where anisotropic stress is
present and approximates the radiative viscosity of a relativistic fluid. The
standard assumption is that $\cvis^2=0$, which however cuts the
Boltzmann hierarchy of perturbations at the quadrupole, forcing a perfect
fluid solution with 
only density, velocity and (isotropic) pressure
perturbations.

Any indication for perturbations in the dark energy fluid would falsify a 
scenario based on the cosmological constant. However, since perturbations 
become observationally unimportant as the 
equation of state approaches the cosmological constant value, $w = -1$, to 
detect them one needs some period in cosmic history when $w$ differs 
substantially from $-1$. Such a deviation in $w$ is constrained at late 
times by the observations, so we are led to consider this at early times, 
along with a non-negligible early dark energy density.

Such early dark energy can arise in some cases of the tracking class of 
dark energy models (see, e.g., \cite{tracking}).  In particular, in 
tracing models the dark energy density is a constant fraction of the 
dominant component, radiation or matter.  If this fraction is non-negligible, 
dark energy could therefore be appreciable not only in the late universe but 
also at early times. Several models of ``early'' dark energy (EDE, hereafter) 
have been proposed (e.g.\ \cite{Doran:2006kp,Linder:2006da} and
references therein). 

Our paper is organized as follows: in Section II we discuss the EDE model 
and the behaviour of perturbations.  In Section III we explain the types 
of CMB data used and the forecast method, including the weak lensing signal. 
In Section IV we present our results, and finally in Section V we discuss 
our conclusions.

\section{Early Dark Energy} 
\label{sec:theory}

\subsection{Model}

\begin{figure}[htb!]
\centering
\includegraphics[width=\columnwidth]{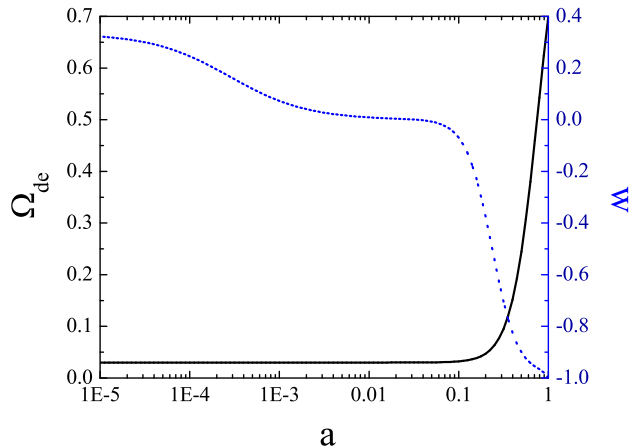}
\caption{ Behaviour of the early dark energy model in energy density (solid
  black line) and equation of state (dotted blue line) as a function of the
  scale factor.}
\label{oe_w}
\end{figure}

In \cite{Doran:2006kp} a parametrization for the dark energy density parameter 
$\od(a)$ and equation of state $w(a)$ has been proposed to recognize the 
important feature of early dark energy.  In this model $\od^0$ and 
$\Omega_{m}^{0}$ are the current dark energy and matter density, 
respectively, and a flat Universe is assumed so $\Omega_{m}^{0}+\od^0 = 1$. 
The model is described by :
\begin{eqnarray}
\od(a) &=&  \frac{\od^0 - \omegae \left(1- a^{-3 w_0}\right) }{\od^0 + \Omega_{m}^{0} a^{3w_0}} + \omegae \left (1- a^{-3 w_0}\right)\\
w(a) &=& -\frac{1}{3[1-\od(a)]} \frac{d\ln\od(a)}{d\ln a} + \frac{a_{eq}}{3(a + a_{eq})}
\end{eqnarray}
where $\omegae$ is the early dark energy component density, constant at high
redshift, $a_{eq}$ is the scale factor at matter-radiation equality, and
$w_0=w(a=1)$.  In Figure~\ref{oe_w} we plot $\od(a)$ and $w(a)$, for $w_0=-1$,
$\omegae=0.03$ and $\od^0=0.7$.  Note the energy density $\od(a)$ goes to a
nonnegligible constant in the past (whereas $\Omega_\Lambda(a=10^{-3})\approx
10^{-9}$).  The dark energy equation of state $w(a)$ clearly shows $3$
different behaviours: $w\sim 1/3$ during the radiation dominated era, $w\sim0$
during matter domination and, finally, $w \sim w_0$ in recent epochs.

Moreover such an EDE model with constant sound speed can behave like
barotropic dark energy models (see e.g. \cite{Linder:2008ya}, and Fig.~2 of
\cite{dePutter:2010vy}).  These models have an explicit relation determining
the pressure as a function of energy density that bring advantages to overcome
the coincidence problem and to predict a value of $w_0\approx-1$ at late
times, considering purely physical properties rather than being adopted as
phenomenology.

Recent analyses have placed constraints on EDE using the available
cosmological datasets and forecasting the discriminatory power of future CMB
probes such as Planck (see
e.g.\ \cite{dePutter:2010vy,Alam:2010tt,Hollenstein:2009ph,Xia:2009ys}).  As
recently shown in particular, the effects of EDE could be important when
combining CMB data with baryonic acoustic oscillation data \cite{linderbao}.
In this paper we follow the lines of these recent papers and we present a
forecast for EDE parameters from the near future Planck \cite{planck} and far
future CMBPol \cite{Bock:2009xw} experiments. Our work will improve similar
recent analyses in several aspects. First, we consider the possibility of
perturbations in EDE including an anisotropic stress term in EDE, parametrized
by a viscosity sound speed $\cvis$ (see \cite{Hu98}). If EDE is following an
equation of state of a relativistic fluid, anisotropic stresses can be present
and change in a substantial way the theoretical predictions on the CMB angular
spectrum. Secondly, we include the CMB weak lensing signal, discussing its
importance in constraining EDE parameters. Finally we also tested our results 
performing a full Monte Carlo Markov Chain (MCMC) on Planck synthetic dataset.

\subsection{Perturbation theory}

Here we briefly review the perturbations in EDE and show theoretical
predictions for the CMB anisotropy angular spectra and for the weak lensing
CMB signal. 

In the synchronous gauge, the energy-momentum conservation in the Fourier
space gives the following equations for the evolution of the density and
velocity perturbations (see \cite{Hu98}, \cite{Ma}) :
\begin{eqnarray}
\label{eq: pert delta}
\frac{\dot{\delta}}{1+w}&=& 
-\left[k^{2}+9\left(\frac{\dot{a}}{a}\right)^{2}\left(\cs^{2}-w+\frac{\dot{w}}{3(1+w)(\dot{a}/a)}\right)\right]\frac{\theta}{k^{2}} \nonumber\\ 
&&-\frac{\dot{h}}{2}-3\frac{\dot{a}}{a}(\cs^{2}-w)\frac{\delta}{1+w} \\
\label{eq: pert theta}
\dot{\theta}&=&-\frac{\dot{a}}{a}(1-3\cs^{2})\,\theta+
\frac{\delta}{1+w}\cs^{2}k^{2}-k^{2}\sigma\\
\label{eq: pert sigma}
\dot{\sigma}&=&-3\frac{\dot{a}}{a}\left[1-\frac{\dot{w}}{3w(1+w)(\dot a/a)}\right]\,\sigma\nonumber\\
& & +\frac{8\cvis^{2}}{3(1+w)}\left[\theta+\frac{\dot{h}}{2}+3\dot{\eta}\right]
\end{eqnarray}
where $\delta$ and $\theta$ are the dark energy density
perturbation and velocity perturbation, $h$ is the metric perturbation 
source, and $-h/2-3\eta$ is the scalar potential of the tensorial metric 
perturbations.  

The above equations describe various models of dark energy; note that 
even if $w(a)$ is the same for two models, they can differ in the 
perturbations.  For a chosen model one can implement these relations 
in a modified version of CAMB
\cite{camb} and solve the Einstein-Boltzmann equations.

\section{Effects on the CMB} \label{sec:cmb} 

\subsection{CMB Angular Spectra}

As already discussed in the literature (see e.g. \cite{Bean} and
\cite{JochenLewis}), perturbations in a dark energy component with a constant
equation of state and a negligible energy component in the early universe
(i.e. $\Omega_e=0$ and $w(a)=w_0$) affect the CMB anisotropy only on very
large angular scales, where cosmic variance dominates.  The reason is that
since in this scenario dark energy contributes appreciable energy density only
at late times and is minimally coupled with other energy components, changes
in the CMB spectra can be only induced by the late Integrated Sachs-Wolfe
(ISW) component.

\begin{figure}[ht!]
\includegraphics[width=\columnwidth]{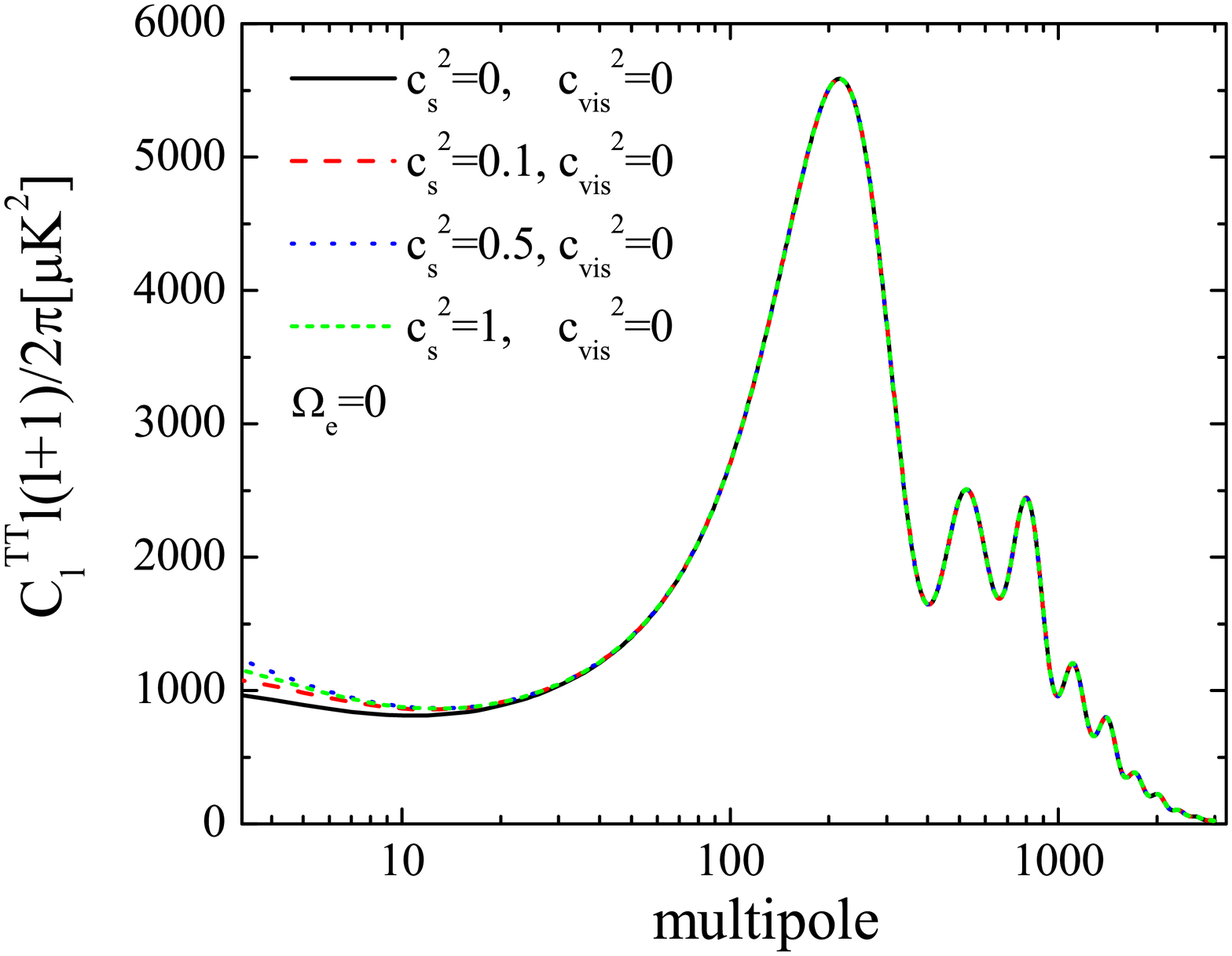}
\includegraphics[width=\columnwidth]{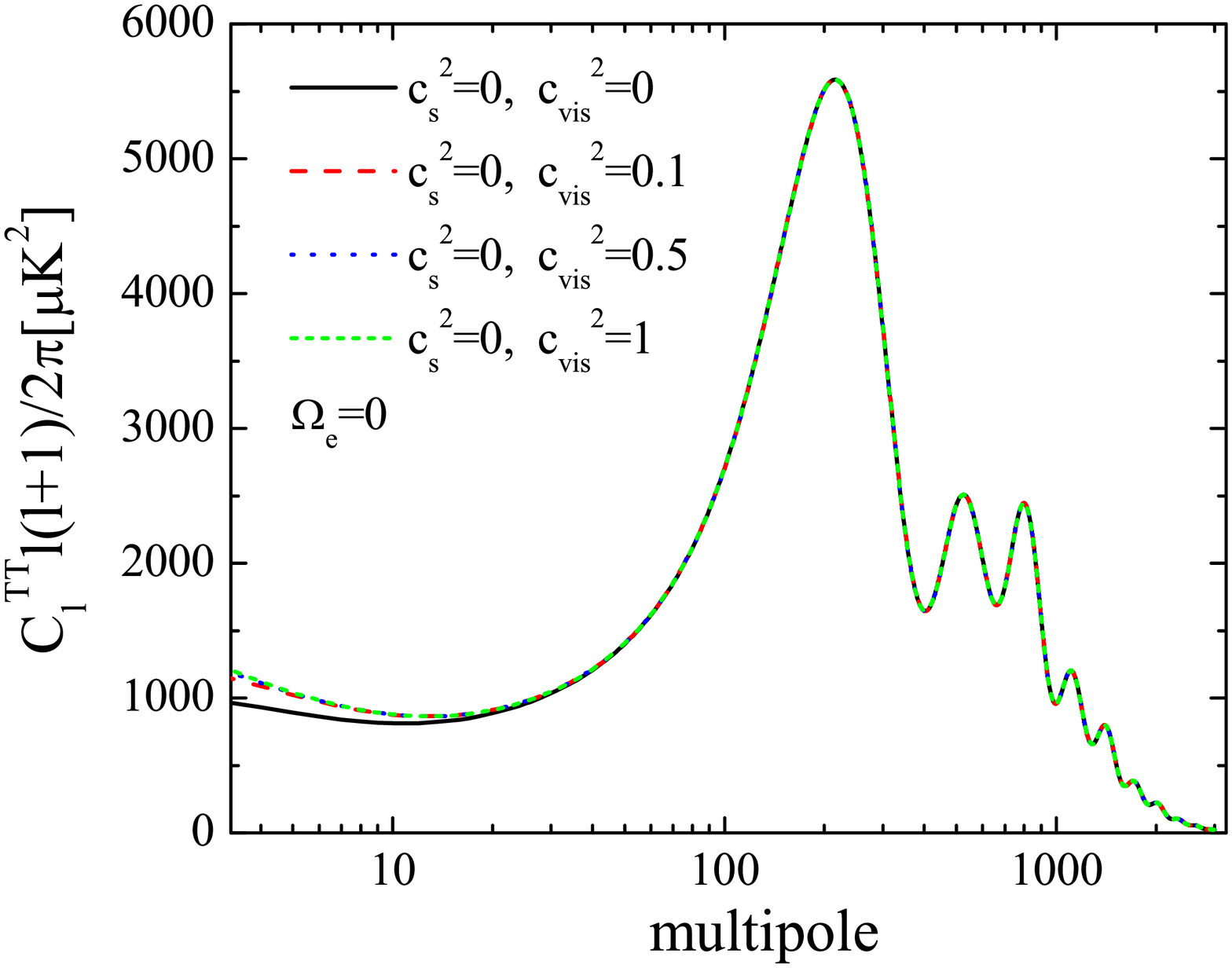}
\caption{Effect of the sound speed (Top Panel) and
viscosity (Bottom Panel) on the CMB spectrum for $\Omega_e=0$
and a constant equation of state $w=-0.8$.}
\label{cvis0.1}
\end{figure}

\begin{figure}[ht!]
\includegraphics[width=\columnwidth]{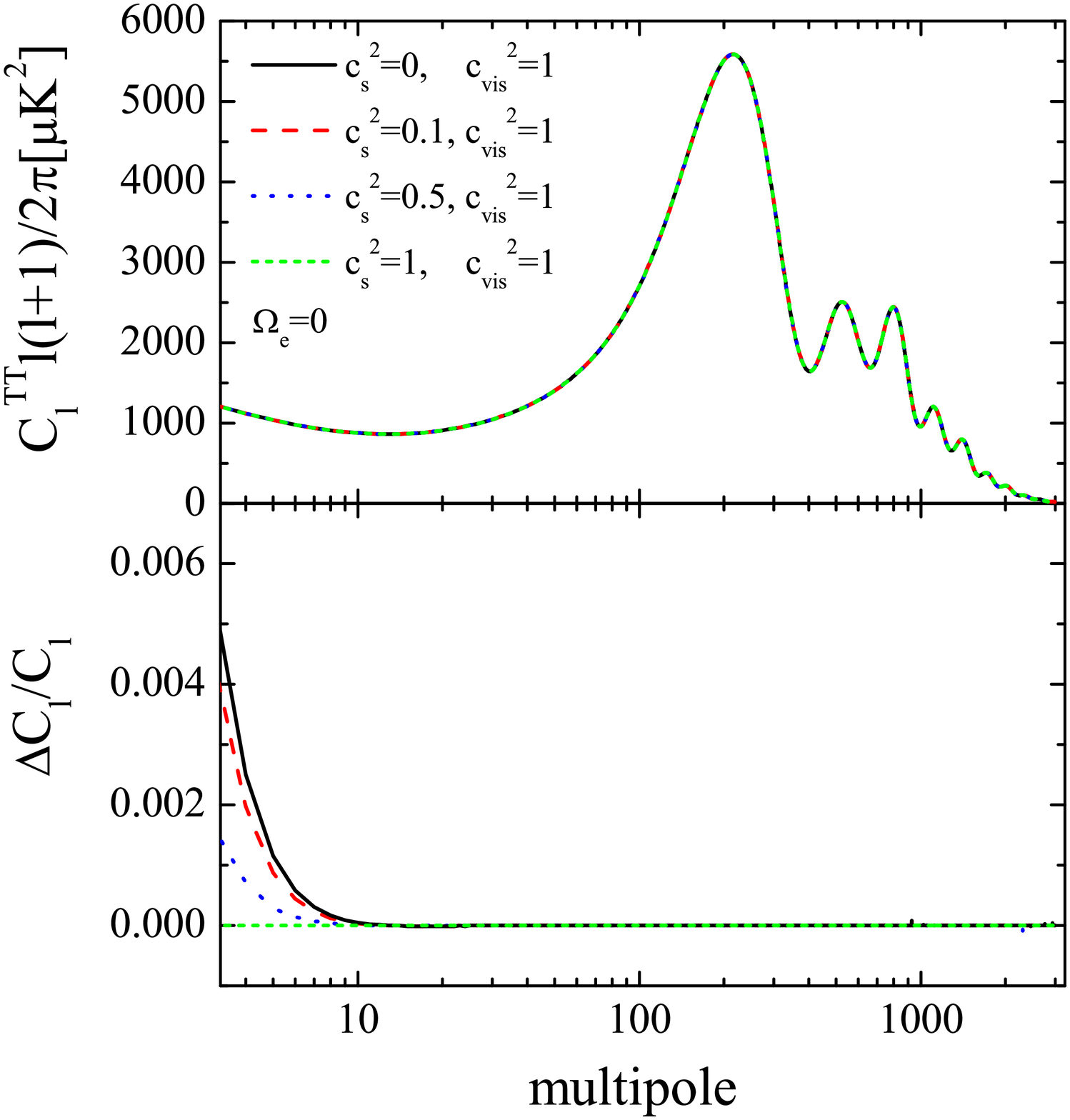}
\includegraphics[width=\columnwidth]{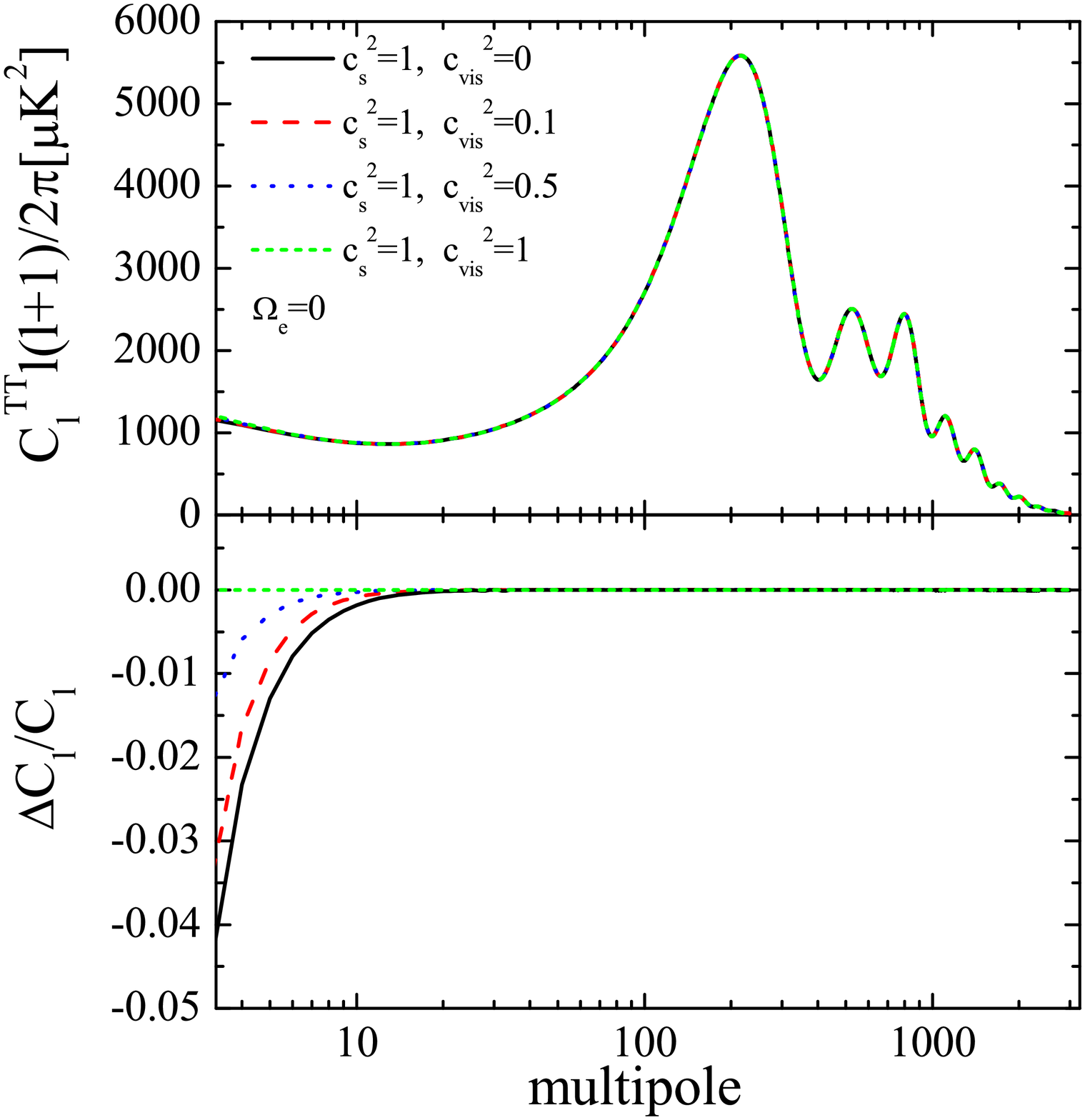}
\caption{As $\cvis^2$ (Top Panel set) or $\cs^2$ (Bottom Panel set) approaches 1, 
the ISW component of the CMB spectrum saturates, bringing essentially 
identical power spectra for different values of the other parameter, i.e.\ 
$\cs^2$ or $\cvis^2$ respectively.  The bottom half of each set shows the 
fractional deviation in power among models.}
\label{cs0.1}
\end{figure}

As an example, we plot in Figure \ref{cvis0.1} the CMB angular spectra for
different values of $\cs^2$ and $\cvis^2$: the variation is only present on
large scales (low multipoles). As already discussed in the literature, the
feasibility of accurately measuring one of these parameters is strongly
undermined by the presence of cosmic variance.  Moreover, the effects of the
two parameters are not uncorrelated with each other, as we show in
Figure~\ref{cs0.1}. Fixing $\cvis^2=1$ or $\cs^2=1$ makes the angular spectra
independent of any variation of the other parameter ($\cs^2$ or $\cvis^2$,
respectively). If one assumes either $\cvis^2=1$ (shown in the top panel), or
$\cs^2=1$ (bottom panel), one is maximally suppressing the perturbations,
giving essentially identical power spectra for different values of $\cs^2$ or
$\cvis^2$, respectively.  This discussion is fully compatible with the results
presented in \cite{mota}.

The net effect of increasing $\cs^{2}$ or $\cvis^{2}$ is higher ISW power. 
This reflects the increased potential decay due to dark energy; while 
dark energy perturbations would help preserve the potential, increasing 
$\cs^2$ or $\cvis^2$ reduces the dark energy perturbation contribution 
and so eases the decay of the potential.  For example, $\Lambda$ leads 
to a high ISW power today.  The effect can be explained more mathematically 
as follows.  The metric perturbation, $h$, is a source term in the
density equation (\ref{eq: pert delta}) and tends to draw dark energy into overdensities of cold dark
matter. 

However, for positive $c_s^2$ and/or positive $c^2_{vis}$, the term proportional
to $\theta$ dominates (on small enough scales) and
suppresses perturbations. In the case of positive $c_s^2$,
this can be seen directly from equation (\ref{eq: pert theta}), where the term proportional to
$c_s^2$ implies that the sign of $\theta$ is the same as that of $\delta$ so that the contribution
to $\dot{\delta}$ has the opposite sign of $\delta$, leading to suppression (a comparison of the magnitudes
of the different terms shows that the suppression becomes dominant roughly
for scales $k > c_s^{-1} \left( \dot{a}/a\right)$).
Thus $\delta$ gets smaller when dark energy begins to
dominate and the ISW effect is enhanced when one increases the sound speed.
In the
case of positive $c^2_{vis}$, it follows from the sign of the metric terms in equation (\ref{eq: pert sigma}) that
$\sigma$ ends up with the same sign as $\delta$, again giving a contribution to the $\dot{\theta}$ equation of the
same sign as $\delta$. Therefore, as the dark energy
becomes dominant, the overall density structure is also smaller when $c^{2}_{vis}$
is larger, and the ISW effect is amplified again.

\begin{figure}[ht!]
\includegraphics[width=\columnwidth]{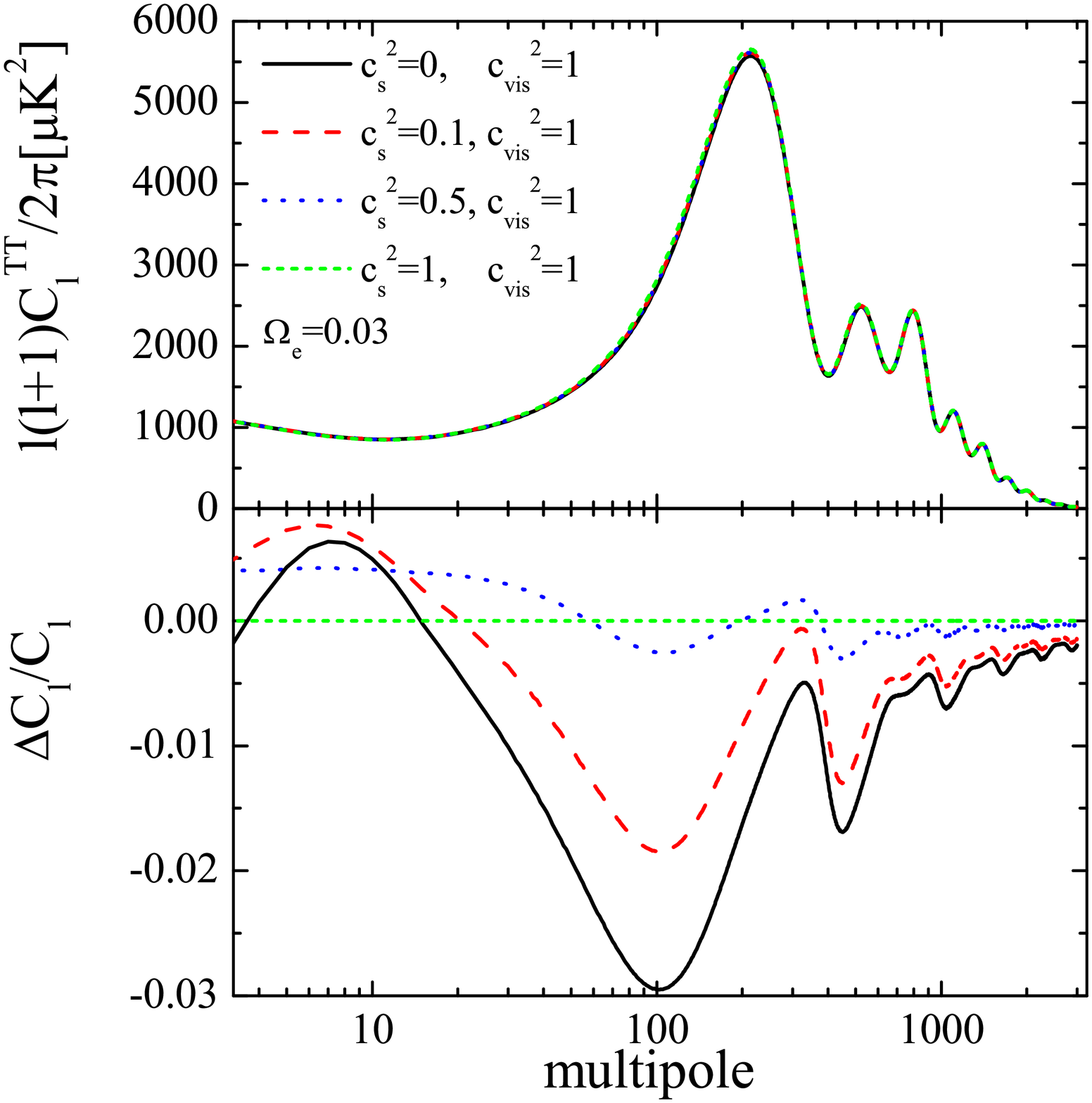}
\includegraphics[width=\columnwidth]{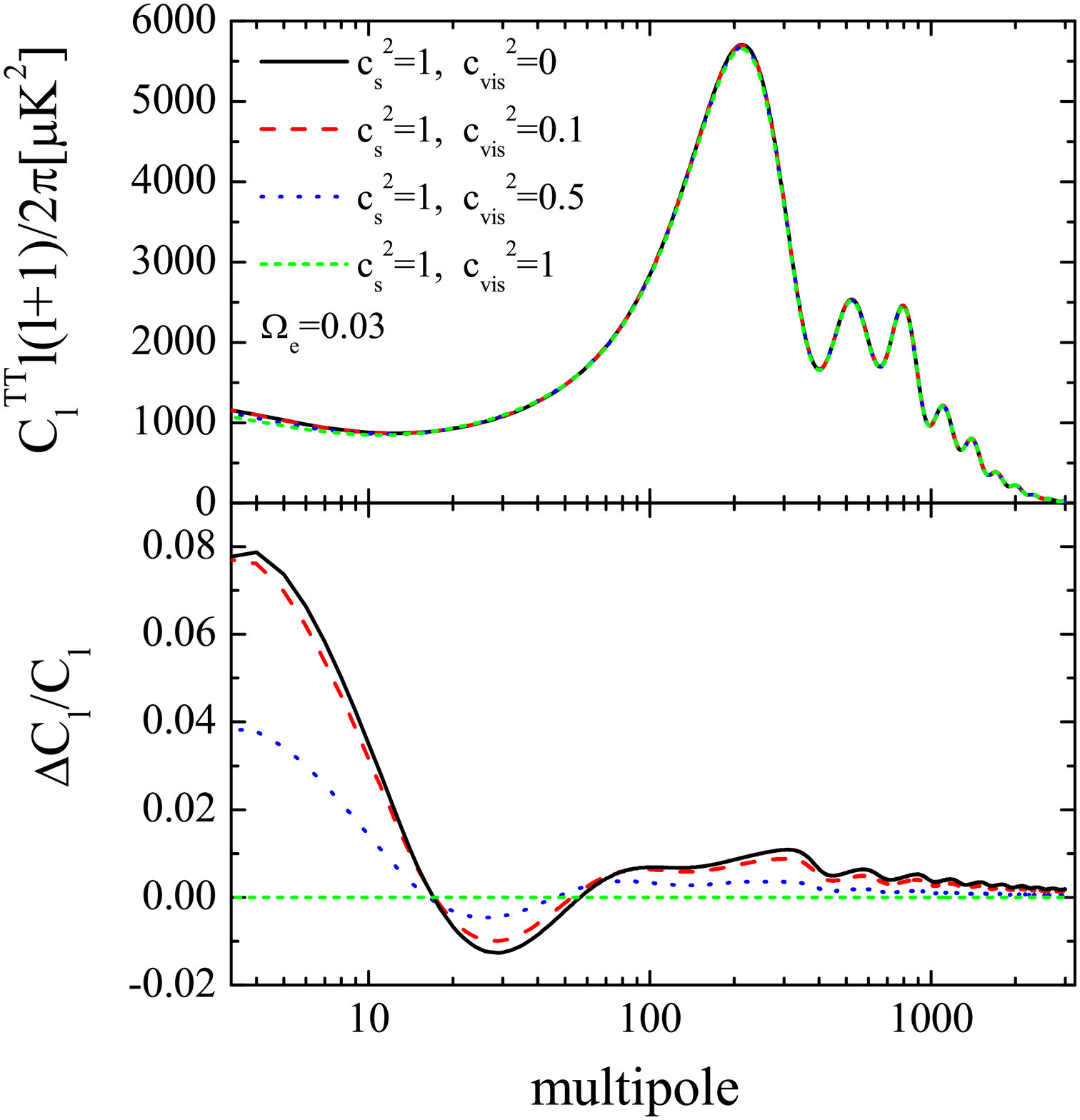}
\caption{Same CMB spectra as in Figure~\ref{cs0.1} but now with an EDE 
component with early energy density $\omegae=0.03$. }
\label{cvis0.1oe}
\end{figure}

It is interesting to investigate if this competition between $\cs^2$ and 
$\cvis^2$ is still present in the case of a EDE scenario. 
For dark energy present at early epochs it may also contribute to the
{\it early} Integrated Sachs Wolfe effect.
In Figure~\ref{cvis0.1oe} we plot the same spectra as in Figure~\ref{cs0.1}
but now with an EDE contribution with $\Omega_e=0.03$.
We see that now the spectra show a small difference around the first peak
due to the different early integrated Sachs Wolfe effect.
While the differences are small it is important to notice that at these
scales the cosmic variance is significantly smaller than at large scales where the 
late-time ISW effect is important.

In addition to the (early and late) ISW effect, the presence of EDE also
affects the evolution of the acoustic oscillations before recombination, leading to
a signature at larger $l$'s than the ISW. If the sound speeds are increased,
EDE perturbations get more suppressed, leading to a stronger decay of the metric
perturbations. This in turn leads to a stronger boost of the amplitude of the acoustic oscillations. 
The (subtle) 
damping in the second peak is a sign that the potentials have not decayed 
as much as when perturbations are unimportant.

\begin{figure}[htb!]
\includegraphics[width=\columnwidth]{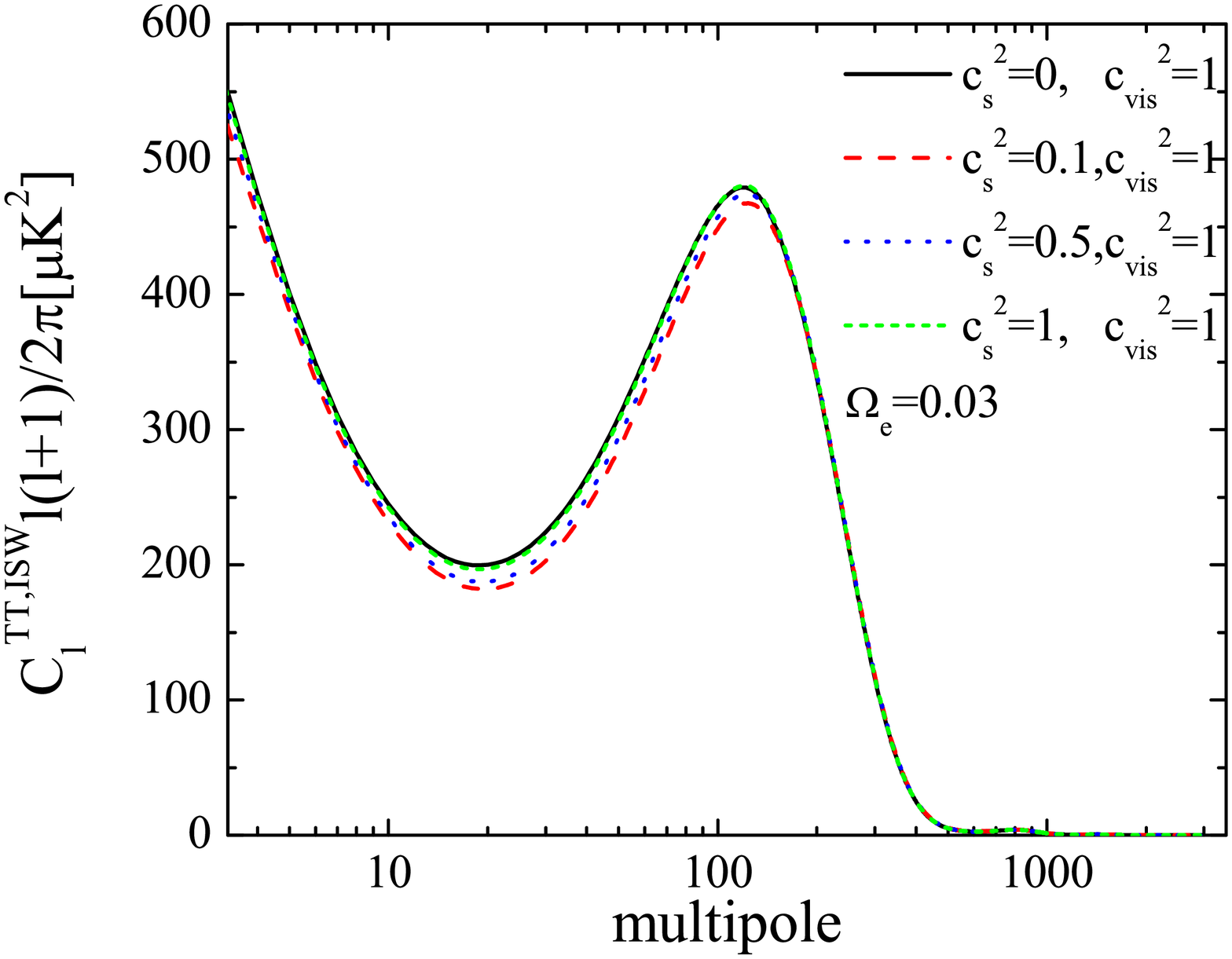}
\includegraphics[width=\columnwidth]{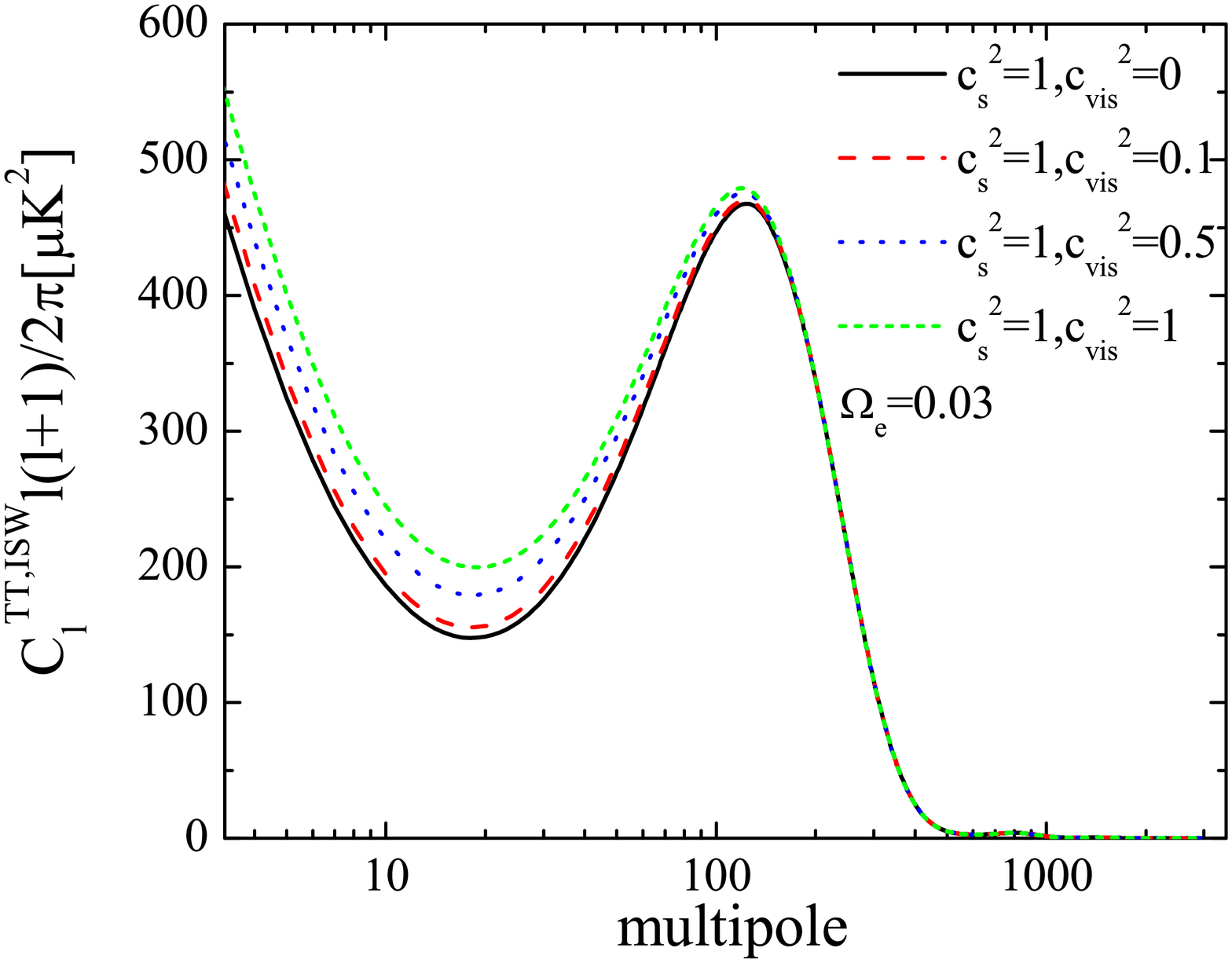}
\caption{As Figure~\ref{cvis0.1oe}, with an early dark energy density 
$\omegae=0.03$, but now focusing on only the ISW component to show the 
effects of the sound speeds. The left rise is due to the 
late ISW effect while the bump is principally coming from 
early ISW. } 
\label{isw_cvis_0.03}
\end{figure}

The ISW behaviour is better shown in Figure~\ref{isw_cvis_0.03} where we plot
just the ISW component of the temperature CMB anisotropy angular power
spectrum.  As we can clearly see, the behaviour of the ISW angular spectrum can be
evidently divided into a contribution from the late ISW effect on large
angular scales ($\ell < 30$) and a contribution from the early ISW, producing a peak on degree
scales at $\ell \sim 120$.  While variations on large scales are negligible
compared to cosmic variance errors, perturbations introduce signal via the early ISW term 
that is more significant.  We
can therefore expect that in the EDE scenario perturbations can play a more
significant role than in a standard late dark energy scenario.  The
perturbations also influence gravitational lensing of the CMB, as we discuss
in the next section.

\subsection{CMB Lensing}

Gravitational lensing of the CMB can improve significantly the CMB constraints
on several cosmological parameters (see e.g.\ 
\cite{Perotto:2006rj,calabrese}), since it is strongly connected with the
growth of perturbations and gravitational potentials. The effect of weak lensing
is to remap the direction of observation (see e.g.\ \cite{IE,Lewis:2006fu,Okamoto2003})
from ${\bf n}$ to ${\bf n}'={\bf n}+{\bf d} ({\bf n})$ 
where ${\bf d}( {\bf n} )$ is the lensing deflection angle.

The lensing deflection angle power spectrum, or equivalently the 
convergence power spectrum, is related to the lensing potential 
spectrum $C_l^{\phi \phi}$, through :
\be
C_l^{dd} = l(l+1)C_l^{\phi \phi} \ .
\ee

\begin{figure}[htb!]
\includegraphics[width=\columnwidth]{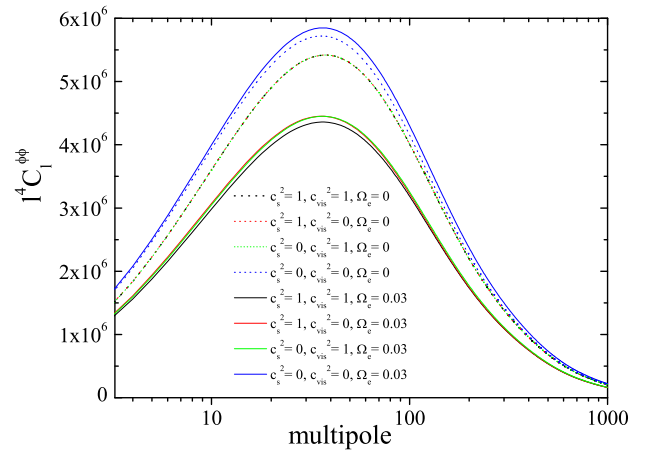}
\caption{Lensing potential power spectra for different standard and early 
dark energy scenarios with $\cs^{2}$ and $\cvis^{2}$ varying from $0$ 
to $1$.}
\label{lens}
\end{figure}

Figure~\ref{lens} shows the lensing potential angular spectra for
scenarios with and without EDE and for different values of $\cs^2$ and
$\cvis^2$.  The plot shows a nontrivial dependence of the lensing angular
spectrum on $\cs^2$, $\cvis^2$, and $\Omega_e$, with some degeneracies clearly
present.  Basically, suppressing perturbations by taking $\cvis^2=1$ or
$\cs^2=1$ (or both) are nearly equivalent.  Only when perturbations are
maximally allowed, through $\cvis^2=0$ and $\cs^2=0$ together, is the lensing power
significantly enhanced.  In this case, early dark energy plays a major role,
yielding a 30\% enhancement in power, while the model with no early dark
energy only sees a $\sim6\%$ boost relative to its no-perturbation case.  We
therefore expect the lensing signal to predominantly improve the constraints
when combined with observations of the primary CMB signal. From
Fig.~\ref{lens} we expect largest improvements on early dark energy, 
but less so on $\cvis^2$ and $\cs^2$, except when they both take on low
values.  We verify this numerically in Sec.~\ref{sec:results}.

\subsection{CMB Experiments and Forecasting} 
\label{sec:fish}

To evaluate the future constraints on EDE models we consider 
the Planck \cite{planck} and CMBPol \cite{Bock:2009xw} experiments using
three frequency channels for each with the experimental specifications as listed in Table~\ref{exp} below.

\begin{table}[!htb]
\begin{center}
\begin{tabular}{rcccc}
Experiment & Channel[GHz] & FWHM & $\sigma_T [\mu K]$ &  $\sigma_P [\mu K]$\\
\hline
Planck & 143 & 7.1'& 6.0 & 11.4\\
$f_{sky}=0.85$
       & 100 & 10.0'& 6.8 & 10.9\\
       & 70 & 14.0'& 12.8 & 18.3\\

\hline
CMBPol & 150 & 5.6' & 0.177 & 0.250\\
$f_{sky}=0.85$ 
       & 100 & 8.4' & 0.151 & 0.214\\
       &  70 & 12.0' & 0.148 & 0.209\\

\end{tabular}
\caption{Planck and CMBPol experimental specifications.}
\label{exp}
\end{center}
\end{table}

We consider for each frequency channel a detector noise of $(\theta\sigma)^2$
where $\theta$ is the FWHM of the beam assuming a Gaussian profile and
$\sigma$ is the sensitivity.
We therefore add to each $C_\ell$ fiducial spectrum a noise
spectrum given by :
\begin{equation}
N_\ell^X = (\theta\sigma_X)^2\,e^{l(l+1)/l_b^2} \, ,
\end{equation}
where $l_b \equiv \sqrt{8\ln2}/\theta$ and the label $X$ refers to either temperature
or polarization, $X=T, P$.

When CMB lensing information is also included we add to our dataset the
lensing deflection angle power spectrum (and the corresponding noise spectrum).  At
sufficiently large angular scales ($l\lesssim 1000$), contributions to the
deflection field will come mainly from the linear regime and, in harmonic
space, the power spectrum of the deflection field reads :
\begin{equation}
\langle a_{lm}^{d*} a_{l'm'}^{d}\rangle =
\left(C_l^{dd}+N_l^{dd}\right) \delta_{ll'}\delta_{mm'},
\end{equation}
where $a_{lm}^d$ can be considered as an approximately Gaussian
variable~\cite{Okamoto2003}.  The noise power spectrum $N_l^{dd}$ reflects the
errors in the deflection map reconstruction. We estimate the lensing
contribution with the quadratic estimator method of Hu \&
Okamoto~\cite{Okamoto2003} based on the correlations between five possible
pairs of maps: $TT$, $EE$, $TE$, $TB$, $EB$ (since the $B$-mode signal is
dominated by lensing on small scales, the estimator $BB$ cannot be used in
this method).  $N_l^{dd}$ corresponds to the minimal noise spectrum achievable
by optimally combining the five quadratic estimators.  Finally, the
non-vanishing correlations between the temperature and the deflection maps are :
\begin{equation}
\langle a_{lm}^{T*} a_{l'm'}^{d}\rangle = C_l^{Td}
\delta_{ll'}\delta_{mm'}~.
\end{equation} 
Following the description in \cite{Perotto:2006rj} we generate $C_l^{dd}$,
$C_l^{Td}$ and $N_l^{dd}$ power spectra and include these datasets in the
analysis, both for Planck and CMBPol. 

To get a general sense of the parameter constraints and degeneracies, 
we first perform a Fisher matrix analysis.  
The Fisher matrix is defined as :
\be
F_{ij}\equiv \Bigl\langle -\frac{\partial^2 
\ln \mathcal{L}}{\partial p_i \partial p_j}\Bigr\rangle _{p_0}
\ee 
where $\mathcal{L}({\rm data}|{\bf{p}})$ is the likelihood function of a set
of parameters ${\bf p}$ given some data; the partial derivatives and the
averaging are evaluated using the fiducial values ${\bf p_{0}}$ of the
parameters.  The Cram\'er-Rao inequality implies that $(F^{-1})_{ii}$ is the
smallest variance in the parameter $p_i$, so 
we can generally think of $F^{-1}$ as the best possible covariance
matrix for estimates of the vector ${\bf p}$. The one sigma error for each
parameter is then defined as :
\begin{equation}\label{sigma}
\sigma_{p_{i}} = \sqrt{(F^{-1})_{i i}}.
\end{equation}

The Fisher matrix for a CMB experiment is given by (see \cite{fishcmb}) :
\begin{equation}
     F^{\rm CMB}_{i j} = \sum_{l=2}^{l_{\rm max}} \sum_{\alpha,\beta}
     \frac{\partial C_l^{\alpha}}{\partial p_i}
     ({\rm Cov}_l)_{\alpha \beta}^{-1}
     \frac{\partial C_l^{\beta}}{\partial p_i},
\label{fishercmb}
\end{equation}
where $\alpha$ and $\beta$ are running indexes over the angular power spectra
$C_{l}$.  For example we include temperature TT, temperature-polarization TE,
E mode polarization EE, or TT, Td, dd in the case
with CMB lensing.  ${\rm Cov}_l$ is the spectra covariance matrix.  We use
information in the power spectra out to $l_{\rm max}=3000$.

\section{Results}
\label{sec:results}
\subsection{Constraints from Planck and CMBPol}

We consider a set of $9$ cosmological parameters with the following fiducial
values: the physical baryonic and cold dark matter densities relative to
critical $\Omega_b h^2=0.02258$ and $\Omega_c h^2=0.1109$, the optical depth
to reionization $\tau=0.088$, the Hubble parameter $H_0=71\, {\rm km/s/Mpc}$,
the current dark energy equation of state $w_0=-0.90$, the early dark energy
density relative to critical $\omegae=0.03$, the spectral index $n_s=0.963$,
and finally the effective and viscous sound speeds
$\cs^{2}$ and $\cvis^2$.  In order to check the stability of the result
under the assumption of the fiducial values for $\cs^2$ and $\cvis^2$ we
investigate several different pairs of values.  CMB lensing is always 
included except for the comparison in Table~\ref{cs_cvis_tab}. 

\begin{figure}[htb!]
\includegraphics[width=\columnwidth]{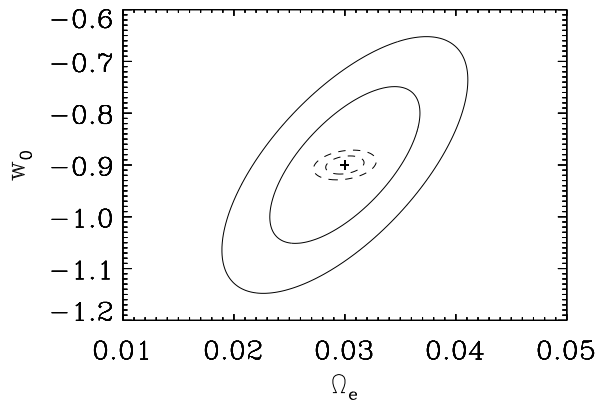} 
\caption{68\% and 95\% c.l.\ likelihood contours for Planck (solid line) 
and CMBPol (dashed line). The `+' symbol represents the fiducial values.} 
\label{oe_wo_cp}
\end{figure}

Using the method described above we forecast the constraints on $w_0$ and
$\Omega_e$.  We find that both Planck and CMBPol can constrain with high 
accuracy those parameters.  Planck will obtain 
$\sigma_{w_0}^{\rm Planck}= 0.10$ while CMBPol can improve this by an order 
of magnitude to $\sigma_{w_0}^{\rm CMBPol}=0.01$.  The density in EDE will 
also be well constrained by Planck, with $\sigma_{\Omega_e}^{Planck}= 0.004$, 
while CMBPol can improve by a factor four 
to $\sigma_{\Omega_e}^{\rm CMBPol}= 0.001$ (see also Table~\ref{tab3}).  
We find no significant dependence of these constraints on the choice of the 
fiducial values of the EDE perturbation parameters $\cs^2$ and $\cvis^2$. 
Figure~\ref{oe_wo_cp} shows the 2-dimensional likelihood plots in the
$w_0$-$\omegae$ plane for both Planck and CMBPol experimental configurations.
These results are for the case $\cvis^{2}= \cs^{2}=0.33$, but again, there
is no practically change in the contours for different choices of $\cs^2$ or
$\cvis^2$.

The expected $1$-$\sigma$ constraints on EDE perturbation parameters 
$\cs^2$ and $\cvis^2$ are presented in Table~\ref{cs_cvis_tab} 
for Planck and for CMBPol experiments.  We show the constraints 
obtained both with and without CMB lensing data.

\begin{table*}[!htb]
\begin{tabular}{||cc||cccc||cccc||}
\hline
& & & & & & & & & \\
& & \multicolumn{4}{c||}{\textbf{No lensing}} & \multicolumn{4}{c||}{\textbf{With lensing}} \\
& & & & & & & & & \\
Fiducial & Fiducial & Planck & CMBPol & Planck & CMBPol & Planck & CMBPol & Planck & CMBPol \\ 
$\cvis^{2}$ & $\cs^{2}$ & $\sigma_{\cvis^{2}}$ & $\sigma_{\cvis^{2}}$ & $\sigma_{\cs^{2}}$ & $\sigma_{\cs^{2}}$ & $\sigma_{\cvis^{2}}$ & $\sigma_{\cvis^{2}}$ & $\sigma_{\cs^{2}}$ & $\sigma_{\cs^{2}}$ \\
& & & & & & & & & \\
\hline\
& & & & & & & & & \\
0.01 & 0.1 & 0.019 & 0.008 & 0.027 & 0.013 & 0.016 & 0.007 & 0.023 & 0.010\\
0.1 & 0.1 & 0.075 & 0.037 & 0.093 & 0.043 & 0.067 & 0.038 & 0.082 & 0.031\\
0.33 & 0.1 & 0.17 & 0.081 & 0.11 & 0.064 & 0.16 & 0.092 & 0.10 & 0.051\\
1 & 0.1 & 0.52 & 0.27 & 0.12 & 0.074 & 0.42 & 0.20 & 0.11 & 0.057\\
& & & & & & & & & \\

0.33 & 0.33 & 0.24 & 0.14 & 0.16 & 0.11 & 0.21 & 0.12 & 0.15 & 0.10\\
& & & & & & & & & \\

0.1 & 0.01 & 0.094 & 0.048 & 0.029 & 0.014 & 0.084 & 0.032 & 0.022 & 0.012\\
0.1 & 0.1 & 0.075 & 0.037 & 0.093 & 0.043 & 0.067 & 0.038 & 0.082 & 0.031\\
0.1 & 0.33 & 0.098 & 0.061 & 0.10 & 0.074 & 0.092 & 0.058 & 0.11 & 0.072\\
0.1 & 1 & 0.19 & 0.10 & 0.71 & 0.35 & 0.17 & 0.091 & 0.68 & 0.33\\

\hline
\end{tabular}
\caption{Fisher analysis results at 68\% c.l.\ for several different 
values of $\cs^2$ and $\cvis^2$, for Planck and for CMBPol datasets, 
with and without CMB lensing included in the analysis. } 
\label{cs_cvis_tab}
\end{table*}

\begin{figure}[htb!]
\includegraphics[width=\columnwidth]{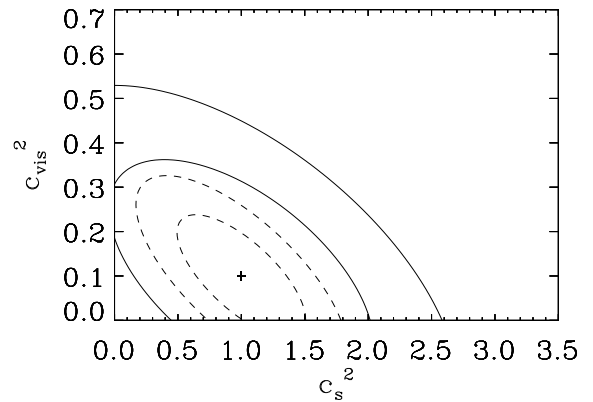}
\includegraphics[width=\columnwidth]{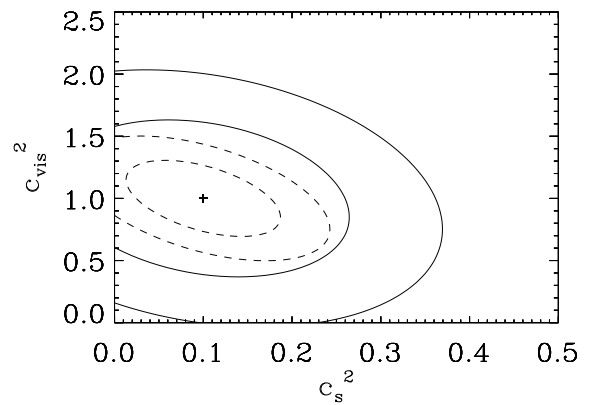} 
\includegraphics[width=\columnwidth]{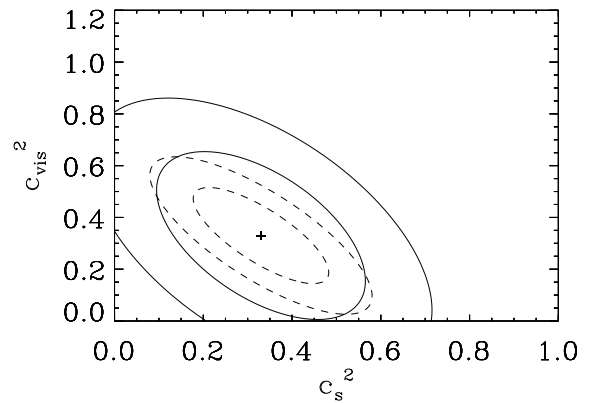} 
\caption{68\% and 95\% c.l.\ likelihood contours for Planck (solid line) 
and CMBPol (dashed line). In the upper panel the fiducial values are 
$\cvis^{2}=0.1$ and $\cs^{2}=1$, in the middle one $\cvis^{2}=1$ 
and $\cs^{2}=0.1$, in the lower panel an intermediate case with 
$\cvis^{2}= \cs^{2}=0.33$ is reported. The `+' symbol represents 
the fiducial values.  Note the different scales.}
\label{cs_cvis}
\end{figure}

From the results listed in Table~\ref{cs_cvis_tab} we can derive the 
following conclusions about estimating $\cvis^2$ and $\cs^2$:

\begin{itemize}
 \item Including CMB lensing improves the constraints by $\sim 10-20 \% $ 
(as compared to $50-60\%$ for $\Omega_e$ and $5-10\%$ for $w_0$). 
 \item  CMBPol provides constraints that are generally a factor $\sim 2$ 
better than Planck.
 \item The constraints on $\cs^2$ (or $\cvis^2$) depend strongly on the 
assumed value of $\cvis^2$ (respectively $\cs^2$), 
the general trend being that
the uncertainties grow with the fiducial values. For example, assuming 
$\cs^2=0.1$, the $1$-$\sigma$ error on this parameter will increase by a 
factor $\sim 5$ if the fiducial model moves from $\cvis^2=0.01$ to 
$\cvis^2=1$. At the same time, assuming $\cvis^2=0.1$, the $1$-$\sigma$ 
error on this parameter will increase by a factor $\sim 2$ if the 
fiducial model moves from $\cs^2=0.01$ to $\cs^2=1$. 
\item The strong correlation between $\cs^2$ and $\cvis^2$ makes it difficult
to precisely measure these parameters individually with either Planck or CMBPol (and 
of course the situation worsens as $\omegae$ decreases or $w_0$ approaches 
$-1$). 
\end{itemize}

The correlation between EDE perturbation parameters can be clearly seen in
Figure \ref{cs_cvis}, where we plot the $68\%$ and $95\%$ c.l. 2-D likelihood
contour plots in the $\cs^2$-$\cvis^2$ plane.  The solid lines are the
constraints derived from Planck while the dashed lines are from CMBPol. 
A reasonable way of quantifying how well the sound speed and viscosity sound speed
can be constrained is by asking at what significance level a non-standard value
of $c^2_s$ or $c^2_{vis}$ can be distinguished from the standard (quintessence) value,
i.e.~from $c^2_s = 1$ or $c^2_{vis}=0$.
By this metric,
whether or not the Planck and CMBpol experiments provide much insight of course depends
on the fiducial values of $c^2_s$ and $c^2_{vis}$.
For example, for the $c^2_s=0.1$ and $c^2_{vis}=1$ fiducial model (middle panel),
Planck could rule out a perfect fluid (i.e.~$c^2_{vis}=0$) at about
$2\sigma$ and CMBpol could do this at more than $4\sigma$. However, for the
$c^2_s=1$ and $c^2_{vis}=0.1$ fiducial model (top panel), neither experiment
can rule out a perfect fluid.
Similarly, for the fiducial in the middle panel, both experiments can rule out
$c_s^2=1$ (quintessence) at very high significance, but not for a fiducial value of
$c_s^2$ significantly closer to unity.

The fact that the uncertainties and ellipse shapes depend strongly on the fiducial parameter values means that
the Fisher matrix evaluated at the fiducial model is not a good predictor of the shape of the likelihood function
away from the fiducial model (and that the likelihood function is thus far from Gaussian). Hence,
away from the fiducial,
the true constant likelihood contours could be quite different from the ones calculated using the Fisher matrix. This means
one has to be cautious when making estimates as in the previous paragraph. For example, from the middle panel
of Figure \ref{cs_cvis}, we estimated that Planck would rule out $c^2_{vis}=0$ at about $2\sigma$, i.e. at $95 \%$
confidence level. However,
since the uncertainty in $c^2_{vis}$ decreases strongly as the fiducial value is lowered, the true significance may
in this case be higher than $95 \%$.
However, this subtlety does not affect the main point made in the previous paragraph, namely that for a range of
reasonable values of $c_s^2$ ($c^2_{vis}$), both Planck and CMBpol will be able to rule out the canonical value,
although CMBpol with much more significance.
In section \ref{sec: MCMC}, we check our Fisher results using an MCMC analysis of
the true non-Gaussian likelihood and we find that our Fisher estimates of uncertainties and error ellipses calculated
are quite accurate.

\subsection{Including Supernovae}

Since the early dark energy component changes the Hubble parameter and 
luminosity distances, Type Ia supernovae (SN) information can be very 
useful to break geometrical degeneracies. 

Each SN magnitude measurement can be expressed as:
\begin{equation}
m_i = 5 \log_{10}[H_0 d_L(z_i, w_0, \Omega_m, \omegae)] + {\mathcal M} 
+ \epsilon_i
\end{equation}
where $d_L$ is the luminosity distance, ${\mathcal M}$ is a combination of the
SN absolute magnitude and Hubble constant, and $\epsilon$ is a zero mean
random term including all systematic and measurement errors.  Given $N$ SN at
redshifts $z_1$...$z_N$, we can describe the measured data $m_i$ as an
N-dimensional vector {\bf m}.  Assuming Gaussian errors $\epsilon_i$, the
Fisher matrix is given by (see \cite{tegmark}) :
\begin{equation}
F^{\rm SN}_{ij} = \frac{1}{2} {\rm Tr} [C^{-1}\frac{\partial C}{\partial p_i} C^{-1}\frac{\partial C}{\partial p_j}] + \frac{\partial {\boldsymbol \mu} ^T}{\partial p_i} C^{-1}\frac{\partial {\boldsymbol \mu}}{\partial p_j} ,
\end{equation}
where ${\boldsymbol {\mu}} \equiv \langle {\boldsymbol m} \rangle$ is the
vector of mean magnitudes and $C \equiv \langle {\boldsymbol m} 
{\boldsymbol m}^T \rangle - {\boldsymbol \mu} {\boldsymbol \mu}^T$ is the 
covariance matrix of magnitudes. The parameter vector ${\boldsymbol p}$ for 
the SN Fisher matrix includes $\Omega_m$, $\omegae$, $w_0$, and the nuisance 
parameter ${\mathcal M}$.

For future SN data we consider 1800 SN out to $z=1.5$ (roughly with a cut 
SNAP distribution \cite{klmm}) plus 300 local ($z=0.05$) SN, with an 
intrinsic dispersion of 0.1 mag and a systematic error
of $0.02(1+z)/2.7$ per $0.1$ bin in $z$ added in quadrature.  New EDE
parameters errors, reported in Table~\ref{tab3}, are estimated considering a
total Fisher matrix :
\begin{equation}
F^{\rm TOT}_{ij}= F^{\rm CMB}_{ij}+F^{\rm SN}_{ij} \ .
\end{equation}

\begin{table}[!htb]
\begin{tabular}{|c|ccc|ccc|}
\hline
& & & & &&\\
Parameter & \multicolumn{3}{c|}{Planck} & \multicolumn{3}{c|}{CMBPol} \\
uncertainty & alone && +SN & alone && +SN \\
\hline
& & & & &&\\
$\sigma_{w_0}$ & 0.10 & & 0.02 & 0.010 & & 0.005 \\
$\sigma_{\omegae}$ & 0.004 & & 0.003 & 0.001 & & 0.001 \\
$\sigma_{\cs^2}$ & 0.15 & & 0.15 & 0.10 & & 0.09 \\
$\sigma_{\cvis^2}$ & 0.21 & & 0.20 & 0.12 & & 0.11 \\ 
& & & &&&\\
\hline
\end{tabular}
\caption{68\% c.l.\ uncertainties on EDE parameters from Planck or CMBPol 
with and without SN distance information.  The fiducial values 
$\cs^2=\cvis^2=0.33$ are used.}
\label{tab3}
\end{table}

We see that the main improvement of adding SNe is on $w_0$, reducing the
Planck uncertainty by a factor of 5, and the CMBPol one by a factor of 2.  The
SN measurements do not reach to high enough redshift to have a good handle on
$\omegae$ (the distance out to $z=2$ in a model with no early dark energy but
$w_a=-(1/2)dw/d\ln a|_{z=1}=5\omegae$ agrees nearly exactly with an EDE model
\cite{linderbao}, and $w_a$ cannot be determined so precisely).  We also see
that the perturbation parameters appear to be mostly uncorrelated with any
parameters to which SN distances are sensitive (indeed, they will be
correlated mostly with each other).  
It is not clear what probes are best for further constraining $\cs^2$ and 
$\cvis^2$, since CMB lensing (especially at the level of CMBPol) already 
includes matter power spectrum information.  Perhaps three-dimensional 
weak lensing and galaxy statistics, or nonlinear structure, would supply 
more leverage.  We leave this for future work.

\subsection{Including Massive Neutrinos}

In addition to considering situations where the perturbation parameter 
constraints improve, we should also explore other parameters that might 
be degenerate with them, and so both weaken the constraints and be 
affected themselves by the presence of cold or stressed dark energy. 

In particular it is interesting to study whether EDE could have any 
implication for the bounds on the neutrino mass from CMB experiments. 
Planck and CMBPol are indeed expected to provide new and very stringent 
bounds on the sum of neutrino masses $\sum m_{\nu}$, extremely 
competitive with respect to bounds coming from laboratory experiments 
as KATRIN \cite{Drexlin:2005zt}.

\begin{figure}[h!]
\includegraphics[width=\columnwidth]{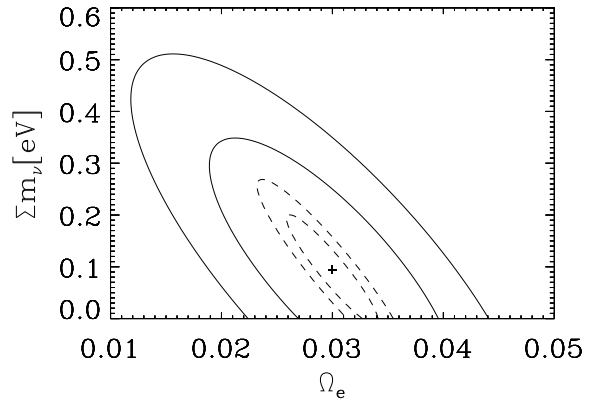}
\caption{68\% and 95\% c.l.\ likelihood contours for Planck (solid line) 
and CMBPol (dashed line).  The degeneracy between $\Omega_e$ and 
$\sum m_\nu$ means that the constraint on $\sum m_\nu$ is 
affected by the inclusion of an EDE component, by a factor 2--3 as seen in 
Table~\ref{tab4}.} 
\label{neutrino-omega}
\end{figure}

We performed a new Fisher matrix analysis adding to our 9-dimensional set of
cosmological parameters the neutrino energy density, $\Omega_{\nu} h^2$, with
a fiducial value of $0.001$ (corresponding to $\sum m_\nu\approx0.09$ eV; 
we quote all results in terms of $\sum m_\nu=94 \Omega_\nu h^2\,{\rm eV}$). 
In Figure~\ref{neutrino-omega} we report the constraints from Planck and
CMBPol and as we can see there is an anticorrelation between $\Omega_e$ and
$\sum m_\nu$ for both Planck and CMBPol experiments. This means that
future CMB bounds on the neutrino mass can be affected by the presence of an
EDE component (also see \cite{dep0901}) . Numerical results are reported in
Table \ref{tab4}. In particular, we studied the impact of one component on the
other. As we can see from the Table the presence of early dark energy and
massive neutrinos almost doubles the uncertainty on both of these parameters.

\begin{table}[!htb]
\begin{tabular}{|c|cc|cc|}
\hline
& & & & \\
Model & \multicolumn{2}{c|}{Planck} & \multicolumn{2}{c|}{CMBPol} \\
& $\sigma_{\omegae}$ & $\sigma_{\sum m_\nu}$ & $\sigma_{\omegae}$ & $\sigma_{\sum m_\nu}$ \\
\hline
& & & & \\
$\omegae=0$ & --  & 0.09 & -- & 0.02 \\
$\sum m_\nu=0$ & 0.004 & -- & 0.001 & -- \\
$\omegae,\sum m_\nu \neq 0$ & 0.007 & 0.20 & 0.003 & 0.07 \\ 
& & & &\\
\hline
\end{tabular}
\caption{68\% c.l.\ uncertainties on EDE density and neutrino density 
from Planck and CMBPol, including marginalization over the perturbation 
parameters $\cs^2$ and $\cvis^2$. 
}
\label{tab4}
\end{table}

Moreover a wrong assumption of the $\omegae$ fiducial value (e.g.\ 
ignoring early dark energy) can bias the estimation of other parameters 
and in particular of neutrino mass, as we now discuss.

\subsection{Bias from Neglecting Perturbations} 

With the Fisher matrix formalism we can also evaluate the bias generated 
in parameter estimation when analyzing the datasets assuming a wrong 
fiducial model, e.g.\ fixing $\cvis^{2}$ to the wrong value. 

For a Gaussian likelihood function, the bias in the $i$-th cosmological
parameter, $\delta\theta_i$, caused by the discrepancy between the
assumed value of a parameter $\psi_j$ and its true value, $\delta\psi_j$, is given by
\cite{Knox_Scocc_Dod,Huterer_Turner,DeBernardis:2008tk} :
\be 
\delta\theta_i = - [F^{\theta\theta}]^{-1}_{ki} F^{\theta\psi}_{k j} \delta\psi_j 
\ee 
where $F^{\theta\theta}$ is the Fisher matrix in the space of $\theta_i$
parameters, and $F^{\theta\psi}$ is a Fisher submatrix with derivatives with
respect to the assumed bias parameters $\psi_j$ and the measured parameters
$\theta_i$. 

\begin{figure*}[htb!]
\centering
\begin{tabular}{cc}
\epsfig{file=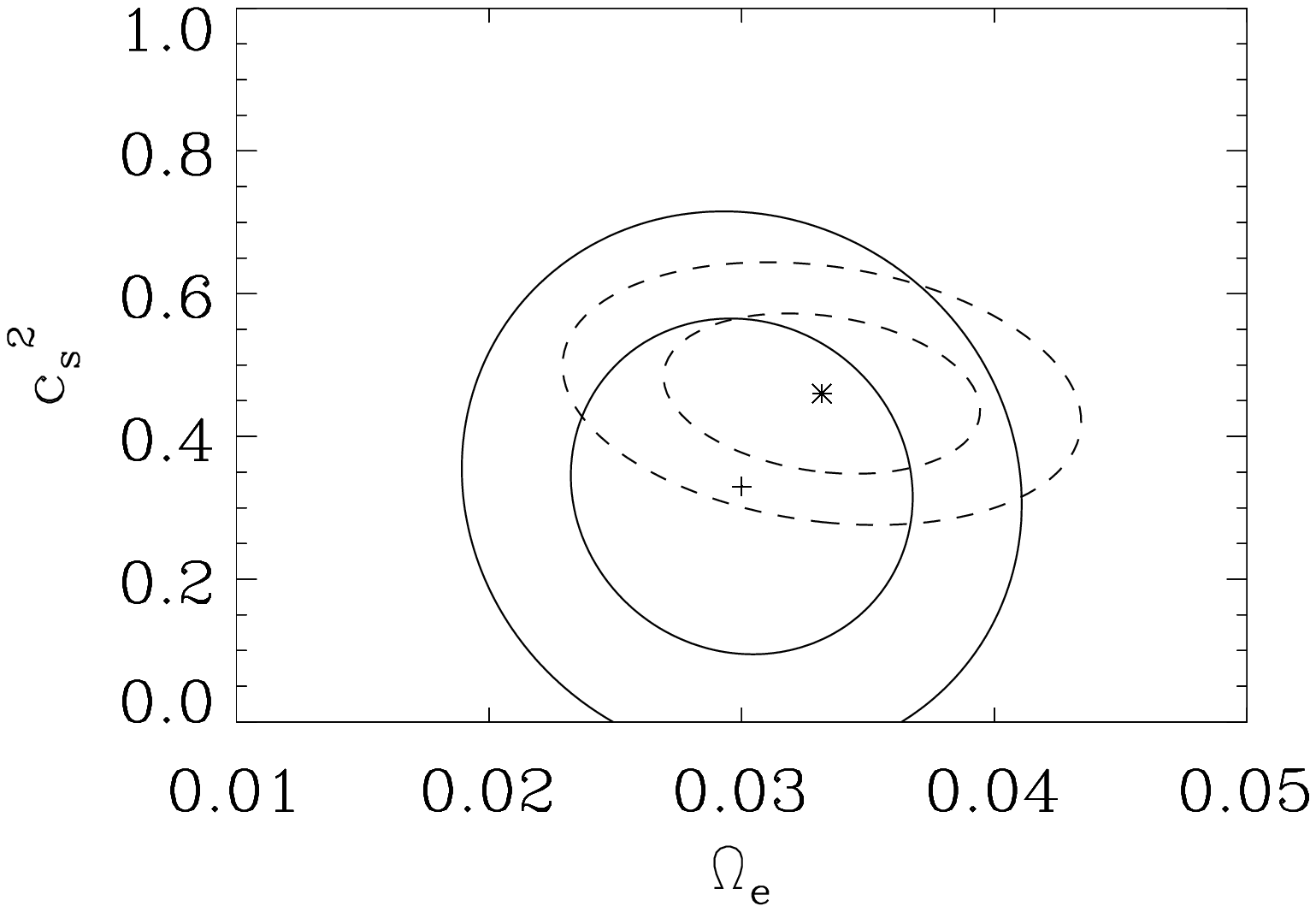,width=0.49\linewidth,clip=} & 
\epsfig{file=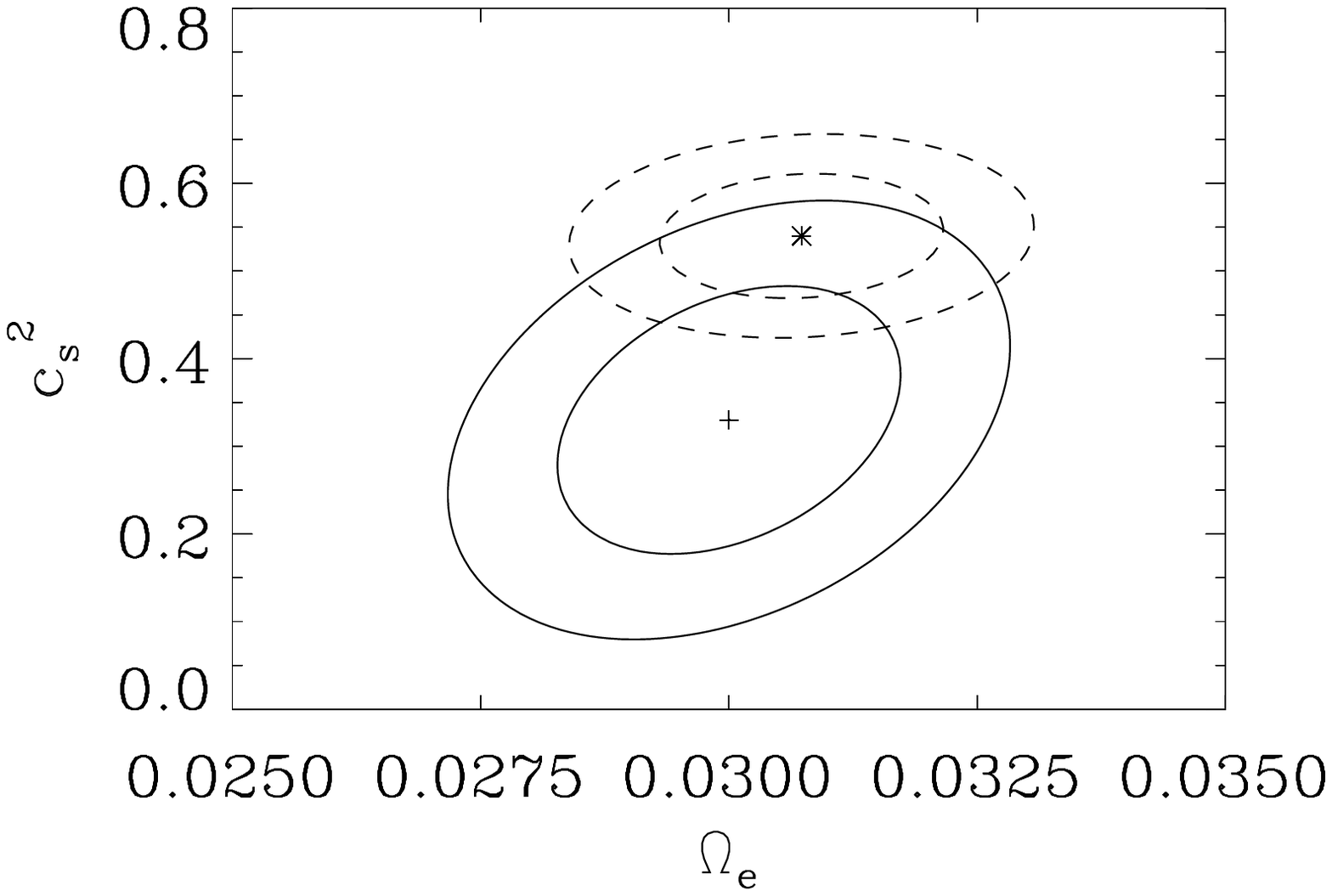,width=0.49\linewidth,clip=} \\
\epsfig{file=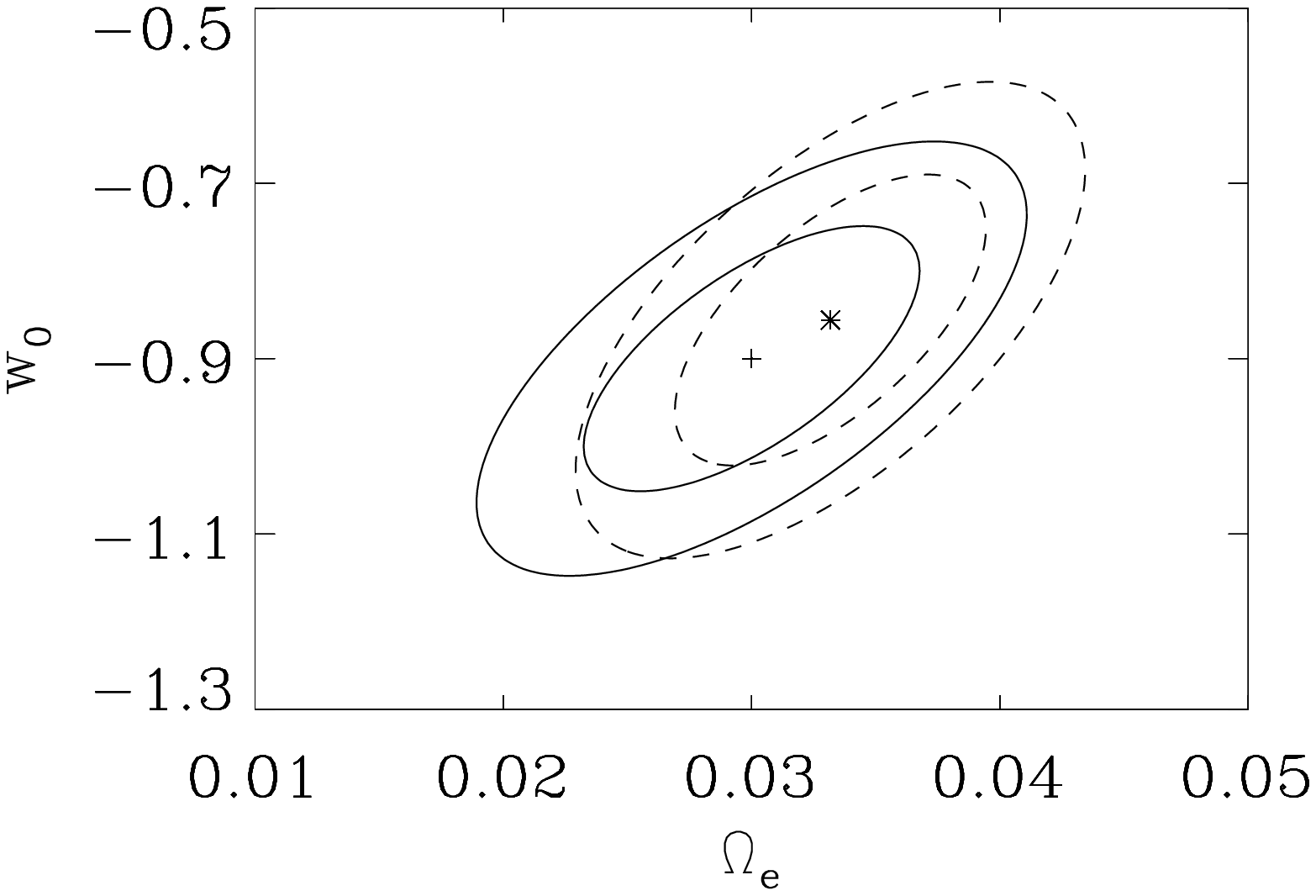,width=0.49\linewidth,clip=} &
\epsfig{file=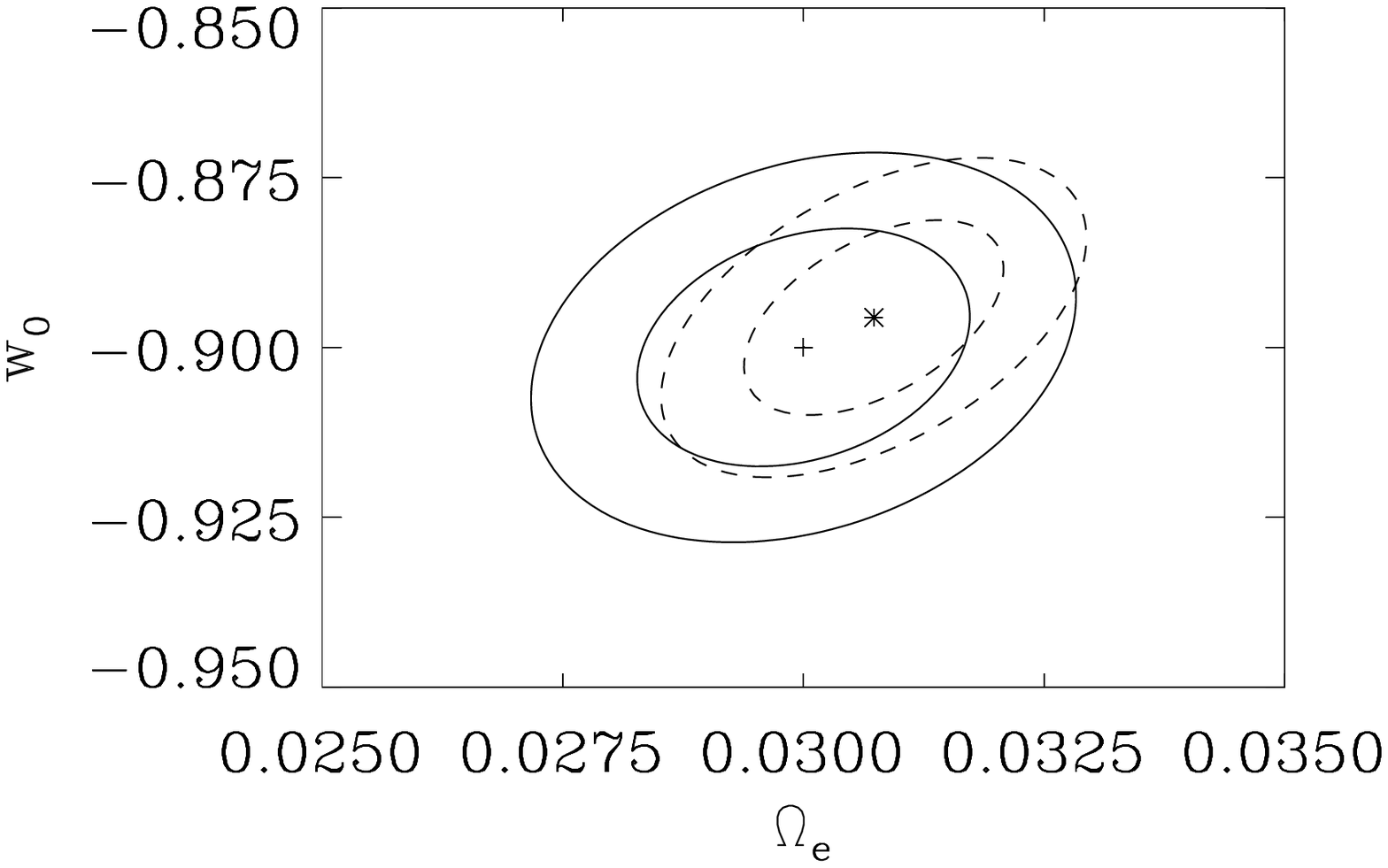,width=0.49\linewidth,clip=}\\
\ \ \ \ \ (a) & \ \ \ \ \ (b)
\end{tabular}
\caption{\label{shift} 68\% and 95\% c.l.\ contour plots in the 
$\omegae$-$\cs^2$ (top panels) and $\Omega_e$-$w_0$ (bottom panels) 
planes for Planck (panel (a)) and CMBPol (panel (b)).  The fiducial 
model has always $\cvis^2=0.33$.  The solid lines are the results 
obtained correctly including $\cvis^2$ in the parameter estimation, 
while the dashed lines are the biased contours obtained when $\cvis^2$ 
is assumed to be 0 (i.e.\ ignoring viscosity).  The `+' symbol 
represents the original fiducial values, while the `$\ast$' symbol gives 
the shifted values.}
\end{figure*}

In our case we want to study the effect of fixing $\cvis^{2}=0$ when an 
input (``true'') model has $\cvis^{2}=0.33$.  Figure~\ref{shift} 
shows the shift obtained on the early dark energy parameters $\cs^{2}$, 
$w_0$, and $\omegae$. We plot 2-dimensional contours showing the 
degeneracies at $68 \%$ and $95 \%$ confidence levels for Planck in the 
left panels and CMBPol in the right panels.  The solid lines are the 
results obtained including $\cvis^2$ in the parameter marginalization, 
while the dashed lines are the contours obtained when $\cvis^2$ is 
(incorrectly) fixed to $0$.

As expected the constraint on $\cs^2$ can be affected by a wrong assumption 
on $\cvis^2$.  Assuming a value of $\cvis^2$ lower than the truth is 
like assuming more perturbations, so $\cs^2$ must be biased high to 
compensate and reduce the perturbations.  The resulting best fit value 
is $\sim1$-$\sigma$ away from the fiducial value for Planck, and 
$\sim2$-$\sigma$ away for CMBPol.   The other parameters are only mildly 
biased. 

When massive neutrinos are considered, $\omegae$ 
will play the major role and will strongly affect $\sum m_\nu$. 
In particular we study the effect of neglecting early dark energy 
(i.e.\ fixing $\omegae=0$) when an 
input true model with $\omegae=0.03$ is used.
This assumption will shift $\sum m_\nu$ from its true value of 
$0.09\,{\rm eV}$ to $0.59\,{\rm eV}$ and $0.65\,{\rm eV}$ for Planck 
and CMBPol respectively -- excluding the true value by $28\sigma$ 
in the latter case!

\subsection{Comparisons with MCMC}
\label{sec: MCMC}

Because the Fisher matrix forecasts are sometimes biased,
especially in case where there is a strong degeneracy between parameters,
we check our previous results with the analysis that maps out the full likelihood 
function in the cosmological parameters.
The analysis uses the publicly
available MCMC package \texttt{cosmomc}
\cite{Lewis:2002ah} with a convergence diagnostic done through the Gelman and
Rubin statistic.  We sample the following 11-dimensional set of
cosmological parameters, adopting flat priors on them: the baryon and cold
dark matter densities $\omega_{\rm b}$ and $\omega_{\rm c}$, the Hubble
constant $H_0$, the scalar spectral index $n_S$, the overall normalization of
the spectrum $A$ at $k=0.05$ {\rm Mpc}$^{-1}$, the optical depth to
reionization, $\tau$, the current equation of state parameter $w_0$, the early
dark energy density $\omegae$, the dark energy sound speed 
$\log\cs^{2}$, the 
viscosity sound speed $\cvis^{2}$, and the neutrino masses $\sum m_\nu$.  We
consider purely adiabatic initial conditions and we impose spatial flatness.
We moreover only consider $w_0$ values greater than $-1$.  
The fiducial model for generating the mock data uses the 
WMAP seven year best fit cosmological parameters values, plus $w_0=-0.9$, $\omegae=0.03$,  
$\cvis^{2}=0.33$ and $\cs^{2}=1$.  

The results obtained are in good agreement with Fisher constraints, 
recovering the fiducial value at the $1\sigma$ level for all the 
parameters. Moreover, the MCMC errors are in good agreement with the 
Fisher matrix error estimates, as reported in Table \ref{mcmc}.

\begin{table}[!htb]
\begin{tabular}{|c|c|c|}
\hline
& &\\
Parameter & Fisher & MCMC \\
\hline
& &\\
$w_0$ & 0.10 & 0.10 \\
$\omegae$ & 0.004 & 0.007 \\
$\cs^2$ & 0.73 & 0.74\\
$\cvis^2$ & 0.26 & 0.27\\ 
& &\\
\hline
\end{tabular}
\caption{1-$\sigma$ errors from Fisher matrix and MCMC analysis on EDE parameters 
from Planck dataset.
}
\label{mcmc}
\end{table}

\section{Discussion}
\label{sec:discussion}

In this paper we have investigated future constraints on EDE models achievable
by Planck and CMBPol experiments.  We included CMB lensing as a probe, 
and the possibilities of a sound speed less than the speed of light and 
of anisotropic stresses in the clustering of the dark energy component
parameterized with a viscosity parameter $\cvis^2$.  Overall, the model 
can be viewed as ``early, cold, or stressed dark energy''.  

We have found that $\cvis^2$ can be strongly correlated with the sound 
speed parameter $\cs^2$. For
this reason it will be difficult for these future experiments to derive
significant constraints on these sound speed parameters individually, 
although finding a deviation from the standard quintessence with
$\cs^2=1$, $\cvis^2=0$ will be possible. 

We have also shown that neglecting the
possibility of anisotropic stresses in EDE could significantly bias the
constraints on EDE parameters. 

The results, obtained through a Fisher Matrix formalism, have been 
checked by a Monte Carlo Markov Chain analysis on Planck synthetic data.  
We have considered SN information to break geometrical degeneracies and 
we have found this significantly improves the equation of state parameter 
estimation. 
Finally we have investigated the impact of EDE on the determination of the
neutrino mass from CMB experiments and we found it to be significant. 
In particular, neglect or misestimation of early dark 
energy density can severely bias neutrino mass constraints for both 
Planck and CMBPol.  
Investigation of early, cold, or stressed dark energy is important 
not only to uncover further windows on the nature of dark energy and 
high energy physics, but to ensure that conclusions on other cosmological 
parameters are robust.

\acknowledgments

DH is supported by the DOE OJI grant under contract DE-FG02-95ER40899, NSF
under contract AST-0807564, and NASA under contract NNX09AC89G.  EL has been
supported in part by the World Class University grant R32-2009-000-10130-0
through the National Research Foundation, Ministry of Education, Science and
Technology of Korea.  RdP, EL have been supported in part by the Director,
Office of Science, Office of High Energy Physics, of the U.S.\ Department of
Energy under Contract No.\ DE-AC02-05CH11231. DH and EL would like to thank
Centro de Ciencias de Benasque ``Pedro Pascual'' for hospitality.


\begin{thebibliography}{}
\bibitem{wmap7}
  E.~Komatsu {\it et al.},
  arXiv:1001.4538 [astro-ph.CO].

\bibitem{acbar}
  C.~L.~Reichardt {\it et al.},
  Astrophys.\ J.\  {\bf 694} (2009) 1200
  [arXiv:0801.1491 [astro-ph]].

\bibitem{quad}
  S.~Gupta {\it et al.}  [QUaD collaboration],
  arXiv:0909.1621 [astro-ph.CO].
\bibitem{SDSS}
  B.~A.~Reid {\it et al.},
  arXiv:0907.1659 [astro-ph.CO].

\bibitem{2dF} W.J. Percival {\em et al.}, \mnras  {\bf 327}, 1297 (2001).

\bibitem{PerlRiess} S. Perlmutter {\em et al.}, \apj {\bf 517}, 565 (1999);\\ 
A.G. Riess {\em et al.}, \apj {\bf 116}, 1009 (1998).

\bibitem{union2} 
R.\ Amanullah et al. 2010, Astrophys. J. 716, 712 [arXiv:1004.1711]

\bibitem{lambda}
  R.~R.~Caldwell and M.~Kamionkowski,
  Ann.\ Rev.\ Nucl.\ Part.\ Sci.\  {\bf 59} (2009) 397
  [arXiv:0903.0866 [astro-ph.CO]];
  P.~J.~E.~Peebles and B.~Ratra,
  Rev.\ Mod.\ Phys.\  {\bf 75} (2003) 559
  [arXiv:astro-ph/0207347];
  E.~J.~Copeland, M.~Sami and S.~Tsujikawa,
  Int.\ J.\ Mod.\ Phys.\  D {\bf 15} (2006) 1753
  [arXiv:hep-th/0603057].

\bibitem{beylam} 
D.\ Rubin et al. 2009, Astrophys. J. 695, 391 [arXiv:0807.1108] 

\bibitem{sollerman} 
J.\ Sollerman et al. 2009, Astrophys. J. 703, 1374 [arXiv:0908.4276] 

\bibitem{mhh2} 
M.J.\ Mortonson, W.\ Hu, D.\ Huterer 2010, Phys. Rev. D 81, 063007 
[arXiv:0912.3816] 

\bibitem{Hu98}
  W.~Hu,
  Astrophys.\ J.\  {\bf 506} (1998) 485
  [arXiv:astro-ph/9801234].

\bibitem{Bean} R.~Bean and O.~Dore, Phys. Rev. D\textbf{69}, 083503 (2004).
\bibitem{JochenLewis} J.~Weller and A.M.~Lewis, \mnras \textbf{346}, 987 (2003).
Astrophys. J. \textbf{617}, L1 (2004). 

\bibitem{Koivisto:2008ig}
  T.~Koivisto and D.~F.~Mota,
  JCAP {\bf 0806} (2008) 018
  [arXiv:0801.3676 [astro-ph]].
\bibitem{mota}
  D.~F.~Mota, J.~R.~Kristiansen, T.~Koivisto and N.~E.~Groeneboom,
  Mon.\ Not.\ Roy.\ Astron.\ Soc.\  {\bf 382} (2007) 793
  [arXiv:0708.0830 [astro-ph]].

\bibitem{tracking}
  I.~Zlatev, L.~M.~Wang and P.~J.~Steinhardt,
  Phys.\ Rev.\ Lett.\  {\bf 82} (1999) 896
  [arXiv:astro-ph/9807002];


\bibitem{Doran:2006kp}
  M.~Doran and G.~Robbers,
  JCAP {\bf 0606} (2006) 026
  [arXiv:astro-ph/0601544].

\bibitem{Linder:2006da}
  E.~V.~Linder,
  Astropart.\ Phys.\ {\bf 26} (2006) 16 
  [arXiv:astro-ph/0603584].
  
\bibitem{Linder:2008ya}
  E.~V.~Linder and R.~J.~Scherrer,
  Phys.\ Rev.\  D {\bf 80} (2009) 023008
  [arXiv:0811.2797 [astro-ph]].

\bibitem{dePutter:2010vy}
  R.~de Putter, D.~Huterer and E.~V.~Linder,
  Phys.\ Rev.\  D {\bf 81} (2010) 103513
  [arXiv:1002.1311 [astro-ph.CO]].

\bibitem{Alam:2010tt}
  U.~Alam, Z.~Lukic and S.~Bhattacharya,
  arXiv:1004.0437 [astro-ph.CO].

\bibitem{Hollenstein:2009ph}
  L.~Hollenstein, D.~Sapone, R.~Crittenden and B.~M.~Schaefer,
  JCAP {\bf 0904} (2009) 012
  [arXiv:0902.1494 [astro-ph.CO]].

\bibitem{Xia:2009ys}
  J.~Q.~Xia and M.~Viel,
  JCAP {\bf 0904} (2009) 002
  [arXiv:0901.0605 [astro-ph.CO]].

\bibitem{linderbao}
  E.~V.~Linder and G.~Robbers,
  JCAP {\bf 0806} (2008) 004
  [arXiv:0803.2877 [astro-ph]].


\bibitem{planck}
    [Planck Collaboration],
  arXiv:astro-ph/0604069.
\bibitem{Bock:2009xw}
  J.~Bock {\it et al.}  [EPIC Collaboration],
  arXiv:0906.1188 [astro-ph.CO].

\bibitem{Ma} C.~P.~Ma and E.~Bertschinger, Astrophys.\ J.\
  {\bf 455} (1995) 7, [arXiv:astro-ph/9506072].

\bibitem{camb}
  A.~Lewis, A.~Challinor and A.~Lasenby,
  Astrophys.\ J.\  {\bf 538} (2000) 473
  [arXiv:astro-ph/9911177].
\bibitem{Perotto:2006rj}
  L.~Perotto, J.~Lesgourgues, S.~Hannestad, H.~Tu and Y.~Y.~Y.~Wong,
  JCAP {\bf 0610} (2006) 013
  [arXiv:astro-ph/0606227].
\bibitem{calabrese}
 E.~Calabrese, A.~Cooray, M.~Martinelli, A.~Melchiorri, L.~Pagano, A.~Slosar and G.~F.~Smoot,
  Phys.\ Rev.\  D {\bf 80} (2009) 103516
  [arXiv:0908.1585 [astro-ph.CO]].

\bibitem{IE}
C.M.~Hirata and U.~Seljak,
Phys.\ Rev.\ D {\bf 68}, 083002 (2003).

\bibitem{Lewis:2006fu}
  A.~Lewis and A.~Challinor,
  Phys.\ Rept.\  {\bf 429}, 1 (2006)
  [arXiv:astro-ph/0601594].



\bibitem{Okamoto2003}
T.~Okamoto and W.~Hu,
Phys.\ Rev.\ D {\bf 67}, 083002 (2003).

\bibitem{fishcmb}
J.~R.~Bond, G.~Efstathiou and M.~Tegmark,
 Mon.\ Not.\ Roy.\ Astron.\ Soc.\  {\bf 291} (1997) L33
 [arXiv:astro-ph/9702100].

\bibitem{tegmark}
  M.~Tegmark, D.~J.~Eisenstein and W.~Hu,
  arXiv:astro-ph/9804168.

\bibitem{klmm} 
A.G.\ Kim, E.V.\ Linder, R.\ Miquel, N.\ Mostek, 
Mon.\ Not.\ Roy.\ Astron.\ Soc.\ {\bf 347}, 909 (2004) 
[arXiv:astro-ph/0304509] 

\bibitem{Drexlin:2005zt}
   G. Drexlin (KATRIN Collaboration), Nucl. Phys. Proc.
  Suppl. {\bf 145}, 263 (2005)

\bibitem{dep0901} 
R.\ de Putter, O.\ Zahn, E.V.\ Linder, 
Phys.\ Rev.\ D {\bf 79}, 065033 (2009) 
[arXiv:0901.0916] 


\bibitem{Knox_Scocc_Dod}
L.\ Knox, R.\ Scoccimarro and S.\ Dodelson, 
Phys.\ Rev.\ Lett.\ {\bf 81}, 2004 (1998) 
  [astro-ph/9805012].

\bibitem{Huterer_Turner}
D.\ Huterer and M.\ Turner, 
Phys. Rev.\ D, {\bf 64}, 123527 (2001) 
[astro-ph/0012510]

\bibitem{DeBernardis:2008tk}
  F.~De Bernardis, R.~Bean, S.~Galli, A.~Melchiorri, J.~I.~Silk and L.~Verde,
  Phys.\ Rev.\  D {\bf 79}, 043503 (2009) 
  [arXiv:0812.3557 [astro-ph]].


\bibitem{Lewis:2002ah}
A. Lewis and S. Bridle,
Phys.\ Rev.\ D {\bf 66}, 103511 (2002) (Available from
\texttt{http://cosmologist.info}.)


\end{thebibliography}
\end{document}